\documentclass[letterpaper,twocolumn,10pt]{article}
\usepackage{usenix-2020-09}

\usepackage{tikz}
\usepackage{amsmath}

\usepackage{filecontents}

\usepackage{booktabs} %
\usepackage{multirow} %
\usepackage{graphicx}
\usepackage{subcaption}
\usepackage{siunitx}
\usepackage[linesnumbered,ruled,vlined]{algorithm2e}
\usepackage{cleveref}

\newcommand{\ours}{\textsf{DVM}}
\newcommand{\eager}{PT-eager}
\newcommand{\tritonr}{Inductor-r}
\newcommand{\tritond}{Inductor-d}
\newcommand{\ms}{MS-O0}
\newcommand{\eagerfull}{PyTorch-eager}
\newcommand{\tritonfull}{TorchInductor}
\newcommand{\tritonrfull}{TorchInductor-recompile}
\newcommand{\tritondfull}{TorchInductor-dynamic}
\newcommand{\msfull}{MindSpore-graph-O0}
\newcommand{\ourspt}{PT-DVM}
\newcommand{\oursms}{MS-DVM}

\begin{document}

\date{}

\title{\Large \bf \ours: A Bytecode Virtual Machine Approach for Dynamic Tensor Computation}

\author{
{\rm Jingzhi Fang}\thanks{These authors contributed equally to this work.}\\
Huawei, China\\
fangjingzhi@huawei.com
\and
{\rm Xiong Gao}\textsuperscript{\hyperlink{footnote.1}{*}}\\
Huawei, China\\
xiong.gao@huawei.com
\and
{\rm Renwei Zhang}\\
Huawei, China\\
zhangrenwei1@huawei.com
\and
{\rm Zichun Ye}\\
Huawei, China\\
zichun.ye@huawei.com
\and
{\rm Lei Chen}\\
HKUST, HKUST(GZ)\\
leichen@cse.ust.hk
\and
{\rm Jie Zhao}\\
Hunan University\\
jiezhao@hnu.edu.cn
\and
{\rm Chengnuo Huang}\\
Huawei, China\\
huangchengnuo1@huawei.com
\and
{\rm Hui Xu}\\
Huawei, China\\
xuhui78@huawei.com
\and
{\rm Xuefeng Jin}\\
Huawei, China\\
jinxuefeng@huawei.com
} %

\maketitle

\begin{abstract}
Dynamism is common in AI computation, e.g., the dynamic tensor shapes and the dynamic control flows in models.
Due to the long compilation time, existing runtime compilation damages the model efficiency, while the offline compilers either suffer from the long compilation time and device memory footprint to cover all the possible execution instances of a dynamic model, or sacrifice optimization opportunities for usability. 
In this paper, we rethink the feasibility of runtime compilation for dynamic models and identify that the key for it to work is to speed up the compilation or hide the compilation overhead. 
To do this, we propose a real-time compiler, \ours.
In \ours, we design a runtime operator compiler based on a bytecode virtual machine to perform effective and efficient compilation for each dynamic operator instance given its input.
Specifically, instead of compiling programs into machine code, we encode the operator program into bytecode on the CPU and decode the bytecode into virtual instructions for direct execution on the NPU.
Based on the runtime operator compiler, we further propose an operator fuser, which performs symbol-deduction-based fusion on static graphs and runtime fusion on dynamic graphs. Both pattern- and stacking-based fusion are supported to increase fusion opportunities.
Evaluation on operators, subgraphs, and models shows that, compared with \tritonfull, \eagerfull~and \msfull, we are up
to {\color{black}11.77$\times$} better in terms of the operator/model efficiency and up to {\color{black}5} orders of magnitude faster in terms of the maximum compilation time.
\end{abstract}

\section{Introduction}

Nowadays, dynamism is common in AI computation. For example, the input sequence lengths of large language models (LLMs) can vary at runtime, making the tensor shapes of operators dynamic, and the control flow in a model also makes the computation graph structure dynamic.
Supporting the dynamic models with dynamic tensor shapes and dynamic topological structures is of significant importance for AI compilers to achieve high model efficiency.

Due to the long compilation time ({\color{black}e.g., according to our experiments, TorchInductor~\cite{torchnpu} can take tens of seconds to compile an operator instance with given input tensors, which can be more than $10^5\times$ of the operator running time}), 
most of the existing compilers for dynamic models depend on offline compilation. 
However, pre-compiled kernels either require a significant overhead of compilation time and device memory footprint to cover all the possible execution instances of a dynamic model in practice, which can often be unacceptable, or sacrifice optimization opportunities for usability.
We elaborate this argument below from two aspects: operator optimization and operator fusion in dynamic models.

\textbf{Operator optimization challenges.}
The naive way to support dynamic operator shapes is to generate kernels for each possible shape, which is impractical when there are a large number of possible tensor shapes, e.g., for the batch inference tasks of LLMs, there can be tens of thousands of possible input shapes $[\mathsf{batch~size}, \mathsf{sequence~length}]$.
Considering that some operator tensor shapes may be more popular than others, just-in-time compilation with caching can be used, but there will be frequent jitter of service time in case of many unseen shapes, and the memory footprint issue remains.

To avoid compiling an operator for every possible shape separately, two types of methods have been developed: (1) bucketing and (2) micro-kernels.
Bucketing~\cite{tvm_bucketing} splits the given possible shape range into several buckets and pads each dimension to the maximum in each range, so that it only needs to generate kernels for the maximum shapes in each range. 
Micro-kernels~\cite{zheng2022dietcode,zheng2023bladedisc,zhou2025sample} are unit computation blocks that can be replicated to constitute a complete operator kernel, i.e., for any operator shape, the complete kernel is determined by the micro-kernel, and different operator shapes can share the same micro-kernel. 
Micro-kernels (and the corresponding kernels) are usually prepared at compile-time, and the best one is selected for the specific shape at runtime.
To obtain the micro-kernel set, some works~\cite{zheng2022dietcode,zheng2023bladedisc} require the possible tensor shape range to define the micro-kernel search space and select a subset of micro-kernels from it based on their performance on the possible shapes. 
When the shape space is large (e.g, an operator has multiple dynamic shape dimensions), it can be hard to search for high-performance micro-kernels efficiently.

Both bucketing and micro-kernels trade off the number of kernels (i.e., the overhead of compilation and memory footprint) against the operator efficiency, and they can fail when the possible shape space is unknown in practice.
Helix~\cite{zhou2025sample} recently constructs micro-kernels offline tailored to the architectural hierarchy directly, requiring no input shape ranges.
However, Helix still suffers from higher operator dynamism, which is the limitation of all the pre-compilation methods.
For example, for the operators with dynamic numbers of dimensions and implicit input broadcast, or the flexibly fused operators (e.g., computation-intensive operators can be fused with element-wise consumers with various operator type combinations), as a kernel cannot be reused among operators with different computation expressions except for different input shapes, it is necessary to prepare a large number of kernels to cover all the possible cases.

\textbf{Operator fusion challenges.}
When fusing dynamic shape operators, existing works rely on symbolic shape equivalence checking~\cite{zheng2023bladedisc}. 
However, when the operator shape equivalence cannot be determined before execution, some fusion opportunities may be missed.
For example, given two operators $A+B, A+C$, supposing the respective shapes of the tensors $A,B,C$ are $[1,20], [b,20], [c, 20]$, the two operators can be fused only when $b=c$ (by fusion, $A$ only needs to be loaded once). If we cannot determine this equality condition beforehand (e.g., the concrete shape is determined by a data-dependent operation like $\mathsf{nonzero}$), the fusion will not be made, even if $b=c$ for some model instances.

The dynamic computation graph structure makes the operator fusion even more challenging.
Existing works~\cite{silvestre2025tempo,zhang2023cocktailer,chen2023dycl} consider three types of control flows in models, i.e., loop, branch, and recursion.
These works either transform the control flow into the data flow (e.g., writing the loops in the kernels or via lifting) or extract sub-graphs without control flows from the computation graph to compile.
However, fusion on complex dynamic topologies has not been well supported yet. 
For example, for the potential fusions across multiple branches, existing works may conservatively not fuse any operators, or we may have to enumerate all the possible operator fusions and prepare respective kernels.
{\color{black}Lazy Tensors~\cite{suhan2021lazytensor} defers operation execution to accumulate a graph and compiles it with the XLA compiler~\cite{xla}, so that the actual execution paths and operator shapes can be utilized.
However, the accumulated graph has to be hashed to avoid unnecessary recompilation, incurring extra overhead.
In PyTorch 2~\cite{ansel2024pytorch}, instead of capturing all the branches of a conditional like a fully symbolic system, it always picks one branch based on shape reasoning given model input and specializes its trace under the assumption that this trace will only be reused when the assumptions hold. 
Both Lazy Tensors and PyTorch 2 suffer from the time-consuming just-in-time recompilation.}

\textbf{Our solution.}
In this paper, we try to solve the problem of dynamic model compilation for better model efficiency and focus on the Ascend NPU architecture. 
Since all of the pre-compilation methods cannot handle the general dynamism by nature, we rethink the feasibility of runtime compilation for dynamic models.
Previous runtime compilation methods cannot work well because we need to (1) either wait for the long compilation to finish before we can run the kernel on the device, (2) or continuously cache the kernels while still suffering from the long compilation when a new operator instance appears.
Therefore, the key for runtime compilation to work is to speed up the compilation or hide the compilation overhead.

To do this, instead of compiling programs to machine code, we turn to a bytecode virtual machine.
Specifically, we encode a program into bytecode on the CPU and decode the bytecode into virtual instructions for direct execution on the NPU.
In this way, we significantly shorten the compilation process compared with the traditional process that generates machine code.
We further speed up the bytecode generation by defining a high-level bytecode per operator type, with each bytecode representing a tile of operator computation (instead of a scalar operation); at runtime, for each fused set of operators, given the exact operator shapes, we use an architecture-guided light-weight shape tiling algorithm to efficiently tile the operator iteration space and generate the corresponding bytecode program.
Each operation bytecode corresponds to a virtual instruction. 
A virtual instruction is executed by calling the corresponding virtual instruction function.
When running a model, thanks to the hardware features of Ascend NPU that (1) different types of computation flows (scalar/vector/matrix computation) run in parallel and (2) the bytecode decoding (requiring scalar computation) is generally much faster than the instruction execution, we can hide the bytecode decoding overhead within the model computation latency.

Based on the runtime operator compiler, we further propose an operator fuser for computation graphs with static or dynamic graph structures.
Two categories of operator fusion, pattern-based fusion and stacking-based fusion, are supported to make the fusion flexible and increase the fusion opportunities.
For static graphs, we check the fusion conditions based on symbol deduction. 
{\color{black}For dynamic graphs, we use an operator buffer to make operator execution lazy and perform streaming operator fusion at runtime based on the actual operator shapes and execution paths, enabling more fusion opportunities than the existing methods.}

\begin{figure}[t]
    \centering
    \includegraphics[width=0.6\linewidth]{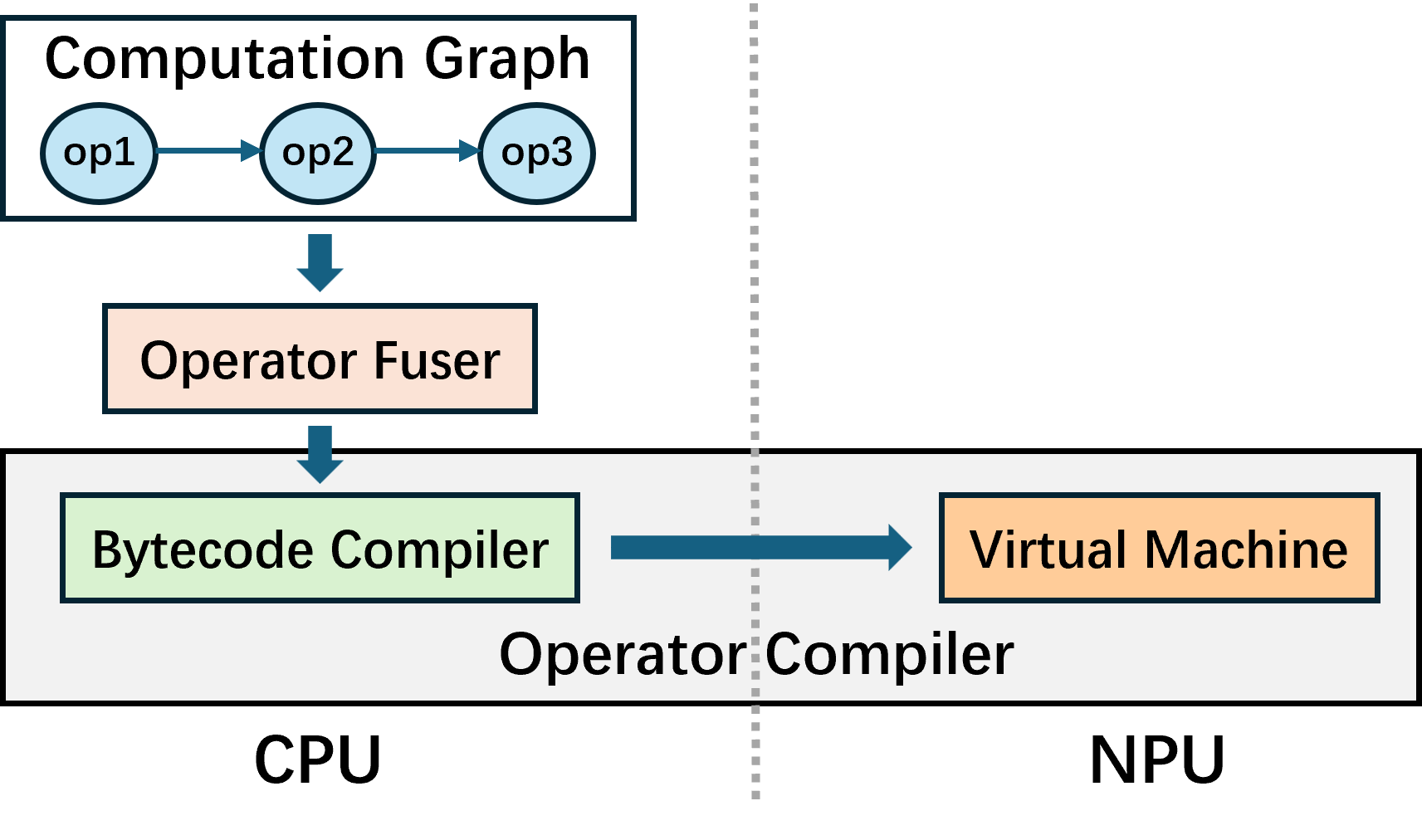}
    \caption{\ours~Overview.}
    \label{fig:dvm overview}
\end{figure}

Building on the above ideas, we propose a dynamism-native compiler, \ours, for dynamic models, which consists of a runtime operator compiler and an operator fuser (\Cref{fig:dvm overview}).
The whole design has 3 advantages.
(1) Dynamism-native: the runtime operator compilation is flexible enough to support various operator shapes and operator fusion.
(2) Light-weight: 
we do not need to enumerate all the possible fused operator kernels before execution, avoiding the high compilation overhead and large memory footprint; 
the compilation is efficient. 
(3) More optimization opportunities: the runtime model information enables more optimization opportunities that are hard to utilize when the operator shape and the execution path are uncertain.
Although \ours~is implemented for Ascend NPU, its design (specifically, the bytecode virtual machine) can be adapted to other accelerators that enable parallel computation flows of different types.

By comparing \ours~with the available Ascend NPU compilers, {\color{black}\tritonfull~\cite{torchnpu} (adapting TorchInductor in PyTorch 2~\cite{ansel2024pytorch} to Ascend NPU), \eagerfull~\cite{torchnpu} (adapting PyTorch eager to Ascend NPU), and \msfull~\cite{mindspore-graph} (the O0 graph mode of MindSpore)} on different operators, subgraphs, and models, we show that \ours~is up to $11.77\times$ better than the baselines in terms of the operator/model efficiency, with the runtime compilation overhead of \ours~counted.
In terms of compilation time, we are up to 5 orders of magnitude faster than the baselines, which shows our significant superiority in rapid model development and deployment.

\begin{figure}[t]
    \centering
    \includegraphics[width=0.8\linewidth]{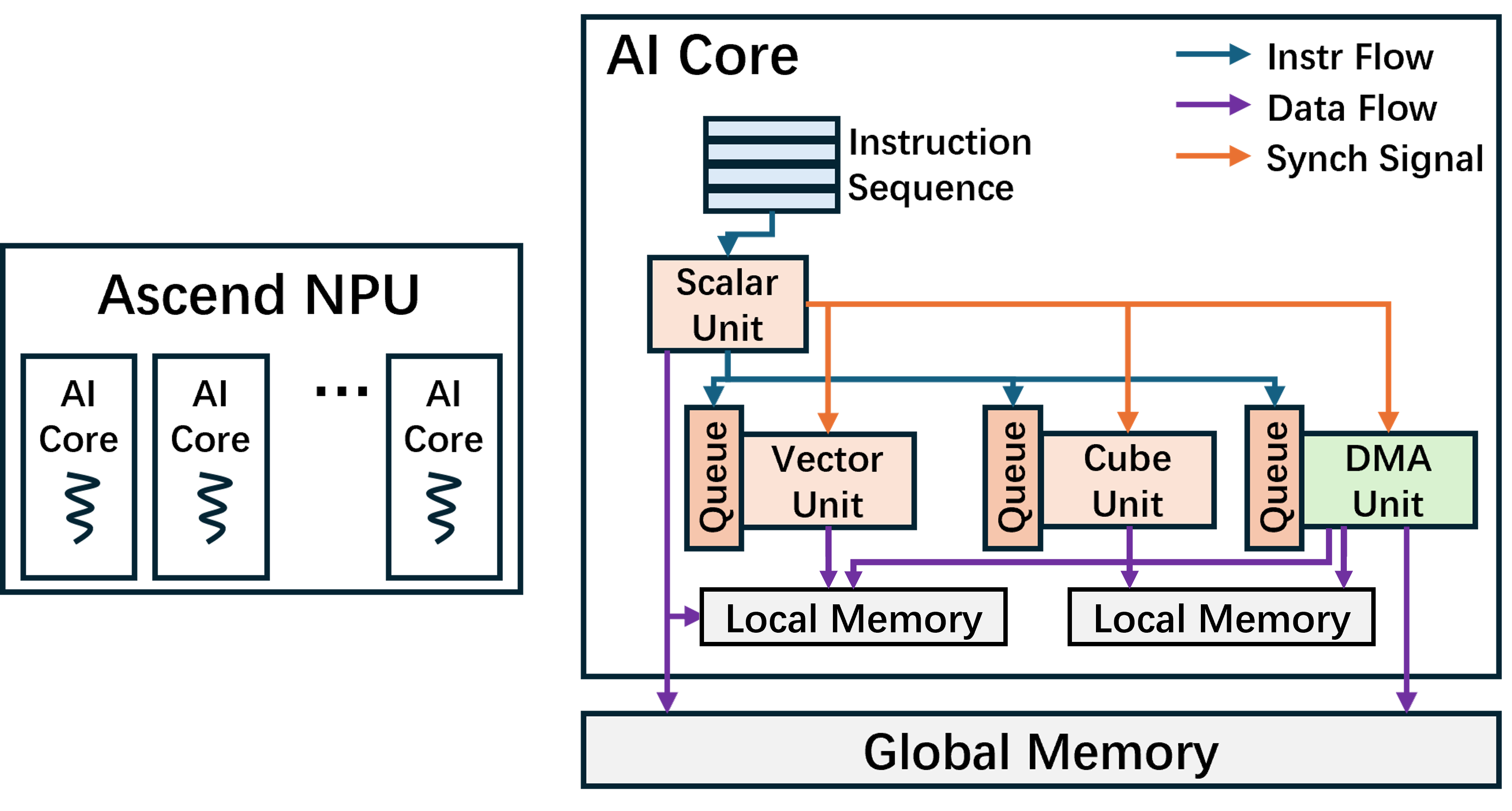}
    \caption{{\color{black}Hardware architecture abstraction (the example AI Core abstraction is about A2/A3 Ascend NPU)~\cite{arch-abstract}.}}
    \label{fig:ai core}
\end{figure}

The contributions of this paper are summarized below.
\begin{enumerate}
    \item We design a runtime operator compiler based on a bytecode virtual machine to achieve both high optimization effectiveness and high compilation efficiency.
    \item We propose an operator fuser based on the runtime operator compiler for both static and dynamic graphs. Particularly, the operator fusion on dynamic graphs is performed at runtime to enable more optimizations. Both pattern- and stacking-based fusion are supported to increase fusion opportunities.
    \item Comparison with the available Ascend NPU compilers, {\color{black}\tritonfull~\cite{torchnpu}, \eagerfull~\cite{torchnpu}, and \msfull~\cite{mindspore-graph}} on different operators, subgraphs, and models shows that we are up to $11.77\times$ better than the baselines in terms of the operator/model efficiency and 5 orders of magnitude faster in terms of the compilation time.
\end{enumerate}

\section{Background}

In this section, we introduce the Ascend NPU architecture and the bytecode virtual machine, respectively.

\begin{figure}[t]
    \centering
    \includegraphics[width=0.9\linewidth]{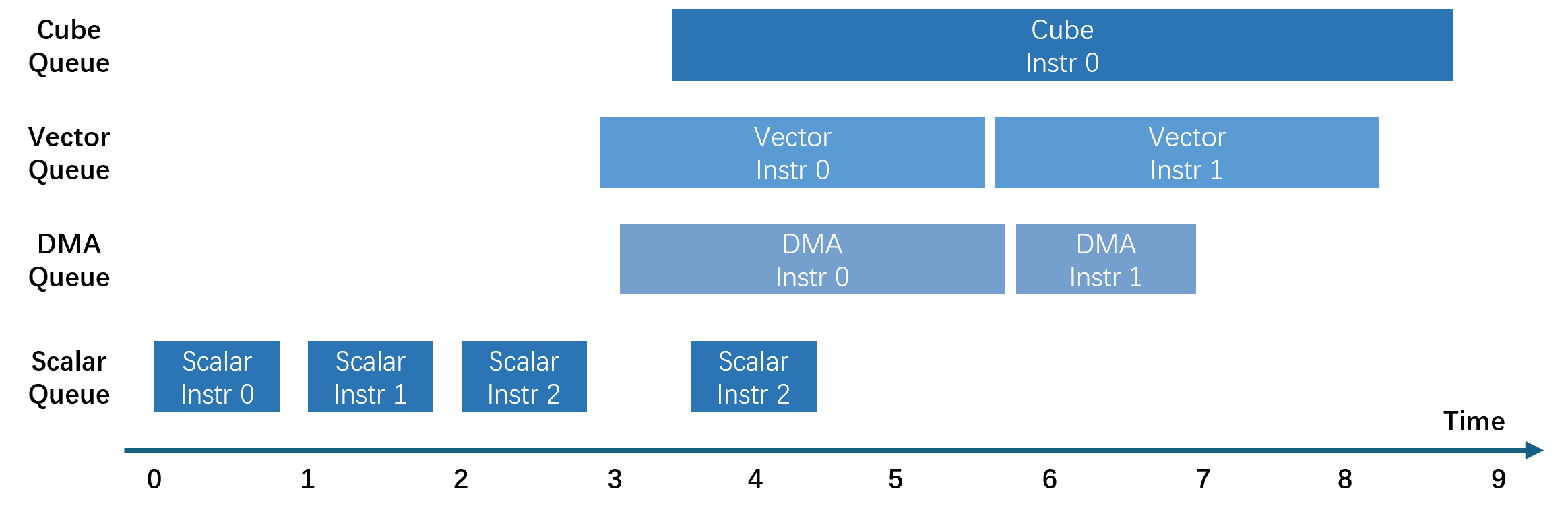}
    \caption{The asynchronous computation flows in an AI Core. }
    \label{fig:comp flow}
\end{figure}

\begin{figure}[t]
    \centering
    \includegraphics[width=0.5\linewidth]{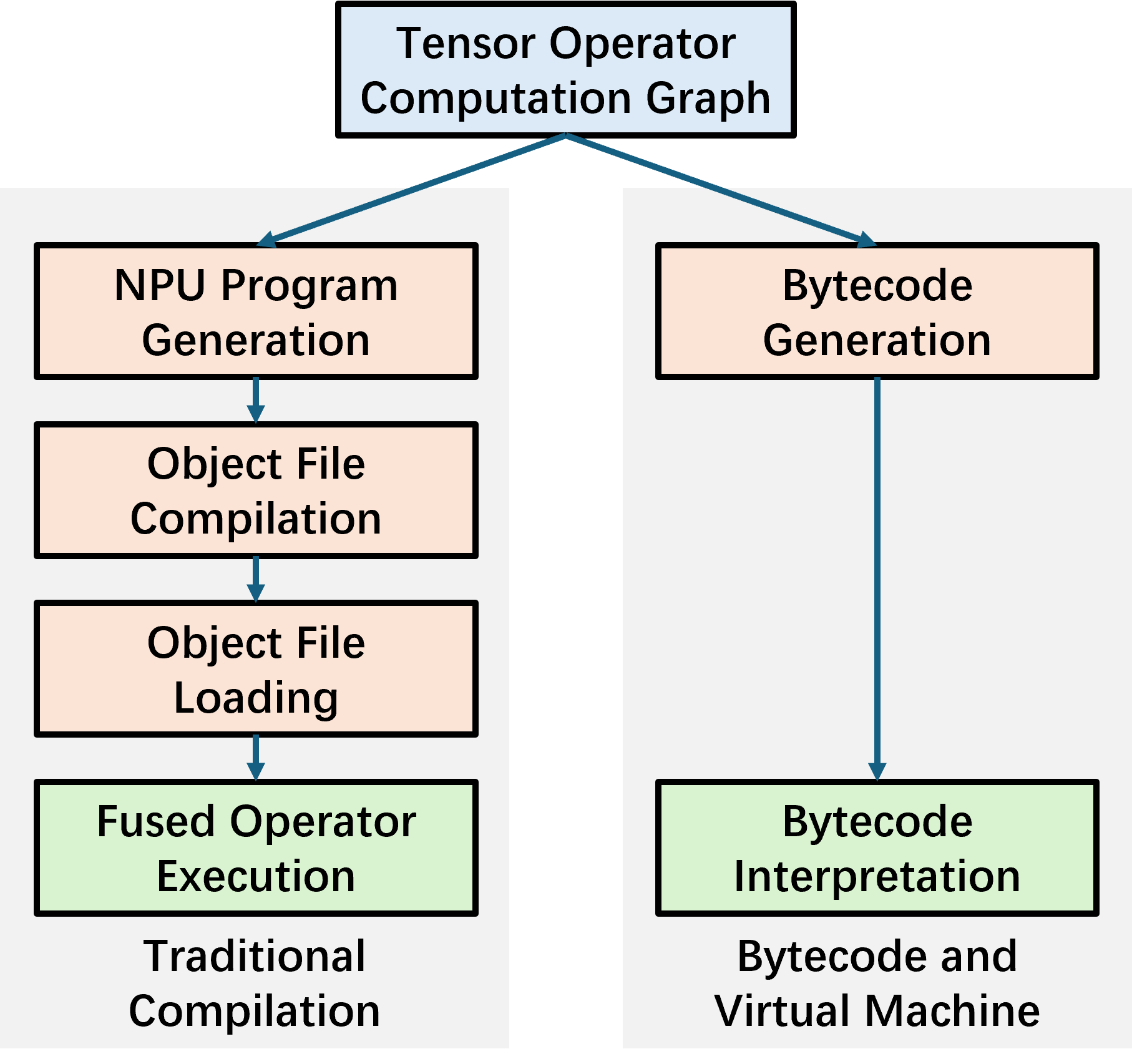}
    \caption{Traditional compilation VS virtual machine.}
    \label{fig:bytecode VM}
\end{figure}

\begin{figure*}[t]
    \centering
    \includegraphics[width=0.9\textwidth]{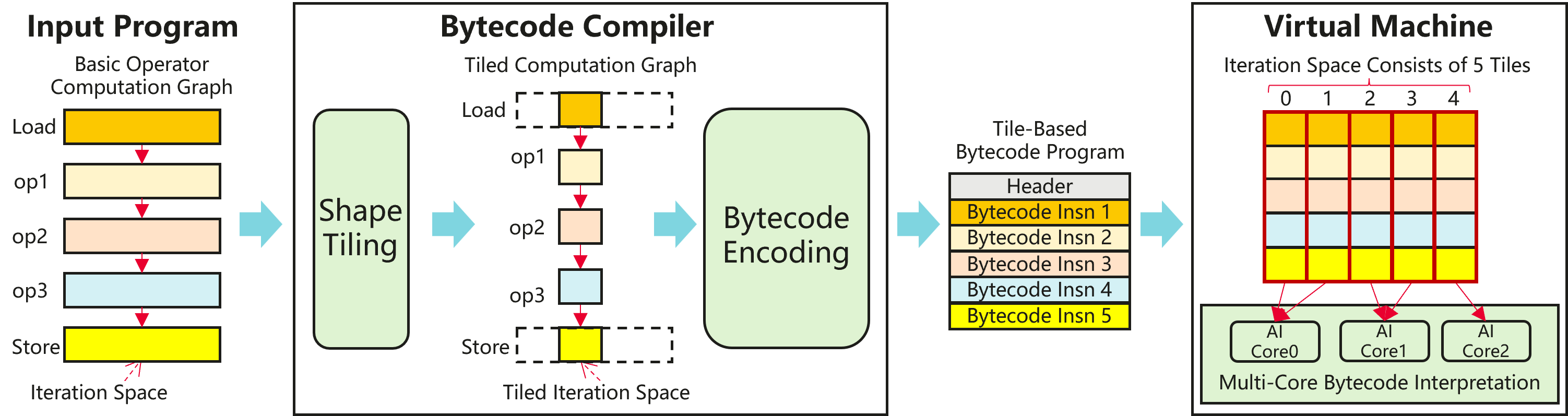}
    \caption{The workflow of the operator compiler.}
    \label{fig:dvm workflow}
\end{figure*}

\subsection{Ascend NPU}
Each Ascend NPU has multiple AI Cores, which can work in parallel.
\Cref{fig:ai core} shows the hardware architecture abstraction of an Ascend NPU~\cite{arch-abstract}.
Each AI core has its computation units, local memory, and {\color{black}the Direct Memory Access (DMA) unit}.
There are three types of computation units: (1) the Scalar unit conducts scalar computation and dispatches the instructions it receives to the corresponding units (the blue arrows in~\Cref{fig:ai core}); (2) the Vector unit conducts vector computation; (3) the Cube unit conducts matrix computation.
The DMA unit is responsible for the data transfer between different memory layers.
The Vector, Cube, and DMA units all have their own instruction queue, so they can run in parallel.
Since there can be dependency between instructions in different queues, the Scalar unit will launch synchronization instructions to the corresponding units when necessary (the orange arrows in~\Cref{fig:ai core}).
A typical dataflow in an AI Core is that DMA first loads data from Global Memory to Local Memory, and then the Vector/Cube unit conducts computation and stores the result in Local Memory for DMA to further move it to Glocal Memory (the purple arrows in~\Cref{fig:ai core}).
\Cref{fig:comp flow} illustrates the asynchronous computation flows of the Ascend NPU.

Ascend NPU conducts computation in an SPMD (Single-Program Multiple-Data) way, i.e., the parallel AI Cores will run the same program on different data, each with one thread.
When calling a function on the Host side (the CPU side), the corresponding task, with the number of AI Cores required and the task type specified, will be loaded by the Device and scheduled to the idle AI Cores, and those AI Cores will run the corresponding kernel function~\cite{hetero-parallel-program}.

\subsection{Bytecode and Virtual Machine}
{\color{black}Unlike compiling a program into machine code to run, a program can also be encoded into a bytecode program, so that a virtual machine can execute it efficiently by decoding it into virtual instructions and directly executing the instructions, one at a time (i.e., performing bytecode interpretation).}
There are bytecode virtual machines for programming languages such as Java~\cite{jvm} and Python~\cite{pvm}, etc.
With bytecode, we do not need to generate or load any file, or compile the program to machine code before execution, i.e., the program execution process (hence the compilation overhead) is significantly shortened compared with the traditional compile-then-execute way (\Cref{fig:bytecode VM}).
Furthermore, since the bytecode is interpreted one at a time, we can pipeline the bytecode decoding and instruction execution to hide the decoding overhead.

\section{\ours~Overview}

We propose \ours, a real-time compiler, which consists of an operator compiler and an operator fuser (\Cref{fig:dvm overview}). 
The operator compiler compiles each (fused) operator instance at runtime, based on bytecode and a virtual machine.
The operator fuser works on both static and dynamic graphs; for dynamic graphs, fusion is based on the actual execution paths and shapes.
Besides, since the dynamic graph structures and the dynamic operator shapes are dealt with at runtime, \ours~does not cache the fusion decisions or the operator kernels, making it memory-efficient.

\textbf{Operator Compiler.}
\Cref{fig:dvm workflow} shows the workflow of our operator compiler.
Like other compilers~\cite{ansel2024pytorch}, the operators in \ours~are decomposed into basic operators for easier handling, e.g., $\mathsf{addmm}$ is decomposed into $\mathsf{matmul}$ and $\mathsf{add}$.
Given a computation graph of a set of basic operators to fuse, the bytecode compiler of \ours~first performs shape tiling to partition the iteration space and then encodes the tiled computation graph into bytecode.
The load/store operations are responsible for the address transformation between the global computation space and the local computation space, so that the diversity of the data layouts in the global computation space is transparent to the tiled computation operations, reducing the complexity of computation operations.
Given the generated bytecode program, the required number of AI Cores for the program will be activated, the computation tiles will be assigned to the AI Cores evenly, and the virtual machine on each AI Core will decode the bytecode program into the corresponding sequence of virtual instructions. Each virtual instruction is executed by calling a pre-compiled kernel function.

We further improve the efficiency of the operator compiler by (1) using a hardware-aligned shape tiling algorithm without real hardware measurement, (2) using tile-level bytecode and virtual instructions to reduce the encoding/interpretation complexity, and (3) pipelining the bytecode generation and interpretation of different fused operators, as well as pipelining the bytecode decoding and instruction execution, to hide the compilation overhead.

\textbf{Operator Fuser.}
The operator fuser groups the operators of the given model to fuse, which will be sent to the operator compiler to compile.
For static graphs with fixed graph structures, we check the fusion conditions, e.g., whether the shapes of two operators can be fused, based on symbol deduction.
For dynamic graphs whose graph structures are uncertain, we perform streaming fusion, so the fusion decisions do not need to be cached, reducing the memory pressure.

\section{Operator Compiler}~\label{sec: dvm}

\begin{table*}[t]
\centering
\caption{Tile-level Virtual Instructions}
\label{tab:virtual_instruction} %
\vspace{6pt} %
\resizebox{\linewidth}{!}{%
\begin{tabular}{cll}
\toprule
\textbf{Category} & \textbf{Instruction}&\textbf{Semantics}\\
\midrule
\multirow{4}{*}{Memory} 
    & \textcolor{black}{Load} (dst, src, tile\_stride, tile\_size) &Load a continuous tile from global memory to local memory\\
    & ViewLoad(dst, src, tile\_stride[], tile\_size[], tile\_dims)  &Load a non-contiguous tile from global memory to local memory\\
    & Store(dst, src, tile\_stride, tile\_size)  &Store a tile from local memory to continuous global memory\\
    & ViewStore(dst, src, tile\_stride[], tile\_size[], tile\_dims)  &Store a tile from local memory to non-continuous global memory\\
\midrule
\multirow{24}{*}{Computation}
    & Copy(xd, xn, size)  &Copy size bytes data from xn to xd in local memory\\
    & Broadcast(xd, xn, M, size, N)  &{\color{black}Broadcast xn of shape [M,1,N] to xd of shape [M, size, N]}\\
    & Sqrt(xd, xn, size)  &xd[i] = sqrt(xn[i])  for i in [0, size)\\
    & Abs(xd, xn, size)  &xd[i] = abs(xn[i])  for i in [0, size)\\
    & Log(xd, xn, size)  &xd[i] = log(xn[i])  for i in [0, size)
\\
    & Exp(xd, xn, size)  &xd[i] = exp(xn[i])  for i in [0, size)
\\
    & Pow(xd, xm, xn, size)  &xd[i] = xm[i]$^{\mathrm{xn[i]}}$  for i in [0, size) \\
    & Round(xd, xn, size)  &xd[i] = round(xn[i])  for i in [0, size)\\
    & Floor(xd, xn, size)  &xd[i] = floor(xn[i])  for i in [0, size)
\\
    & IsFinite(xd, xn, size)  &Check finiteness of each xn[i], store to xd[i], for i in [0, size)\\
    & Adds(xd, scalar, size)  &xd[i] = xd[i] + scalar  for i in [0, size)
\\
    & Muls(xd, scalar, size)  &xd[i] = xd[i] * scalar  for i in [0, size)
\\
    & Add(xd, xm, xn, size)  &xd[i] = xm[i] + xn[i]  for i in [0, size)
\\
    & Sub(xd, xm, xn, size)  &xd[i] = xm[i] - xn[i]  for i in [0, size)
\\
    & Mul(xd, xm, xn, size)  &xd[i] = xm[i] * xn[i]  for i in [0, size)
\\
    & Div(xd, xm, xn, size)  &xd[i] = xm[i] / xn[i]  for i in [0, size)
\\
    & Min(xd, xm, xn, size)  &xd[i] = min(xm[i], xn[i])  for i in [0, size)
\\
    & Max(xd, xm, xn, size)  &xd[i] = max(xm[i], xn[i])  for i in [0, size)
\\
    & \multirow{2}{*}{Cmp(xd, xm, xn, size, cmp\_type)}  & xd[i] = (xm[i] <cmp\_type> xn[i]) ? 1 : 0  for i in [0, size) \\
    &                                   & cmp\_type: EQ, NE, LT, LE, GT, GE \\
    & Cast(xd, xn, size, src\_dtype, dst\_dtype)  & xd[i] = (dst\_dtype)(xn[i])  for i in [0, size) \\
    & Sum (xd, xn, M, size, N)  & {\color{black} sum xn of shape [M, size, N] along the penultimate dimension to xd of shape [M, 1, N]} \\
    & Max(xd, xn, M, size, N)  & {\color{black} compute max on xn of shape [M, size, N] along the penultimate dimension and store it in xd} \\
    & Min(xd, xn, M, size, N)  & {\color{black} compute min on xn of shape [M, size, N] along the penultimate dimension and store it in xd} \\
    & Select(xd, cond, xm, xn, size)  &  xd[i] = (cond[i] != 0) ? xm[i] : xn[i]  for i in [0, size) \\
\bottomrule
\end{tabular}
}
\end{table*}

Given an operator (or a set of operators), the bytecode compiler encodes it into bytecode on the host, and the virtual machine interprets the bytecode on the device to perform the actual computation.
In this section, we first introduce the virtual instructions used by \ours, and then present the details of the bytecode compiler and the virtual machine, respectively.

\subsection{Virtual Instruction}

Each virtual instruction in \ours~corresponds to a kernel function on the device that performs a tile of a certain computation.
For example, $\mathsf{Add(xd, xm, xn, size)}$ computes the addition of two input tensors $\mathsf{A[size]},\mathsf{B[size]}$ with addresses $\mathsf{xm, xn}$ and stores the results in memory with the address $\mathsf{xd}$.
Compared with the traditional scalar virtual instructions, such tile-level virtual instructions have two advantages:
(1) fewer instructions are required for the same program, so the time of bytecode encoding and decoding can be shorter;
(2) the computation tiles are aligned with the SIMD (Single-Instruction Multiple-Data) architecture of the Ascend NPU, so the computation resources can be fully utilized and the running efficiency can be higher.

\Cref{tab:virtual_instruction}~lists the virtual instructions used by \ours.
Specifically, we split the virtual instructions into two categories: memory-related (load and store) and computation-related virtual instructions.
In \ours, we ensure that the data layout in the local memory is the same as the computation-related virtual instructions expect, so the memory-related virtual instructions will automatically transform the data layout when moving data between the global and the local memory.
In this way, we do not need to design different computation-related virtual instructions for different data layouts, which simplifies the implementation and reduces the size of the virtual instruction set.

\begin{algorithm}[t]
\caption{Shape Tiling}
\label{alg: shape tiling}
\KwIn{Basic operator computation graph $G$, the local memory size $M$}
\KwOut{Tiled basic operator computation graph $G'$}
$S_d \leftarrow$ the dominant shape of $G$\;
$n_{\mathsf{max}} \leftarrow$ the peak number of live operators in $G$\;
$L \leftarrow$ the total size of $S_d$\;
$T_{\mathsf{max}} \leftarrow M/n_{\mathsf{max}}$; {\small \color{olive}// the tile size limit}

$i\leftarrow$ the outermost dimension of $S_d$\;
$G'\leftarrow G$\;
\While{ $L>T_{\mathsf{max}}$ }{
    $\ell \leftarrow S_d[i]$\;
    \If{$L/\ell > T_{\mathsf{max}}$}{
        $t\leftarrow 1$\;
        $L\leftarrow L/\ell$\;
    }
    \Else{
        $L'\leftarrow L/\ell$\;
        $t\leftarrow \mathsf{hardware\_align\_div(T_{\mathsf{max}}, L')}$\;
        $L\leftarrow t*L'$\;
    }
    \For{$\mathsf{op} \in G'$}{
        {{\small \color{olive}// skip broadcasting or reduction dimensions}}
        
        \If{$\mathsf{op.shape[dim]}\ne 1$ }{
            $\mathsf{op.shape[dim]} = t$\;
        }
    }
    {$i\leftarrow$ the next inner dimension of $S_d$}\;
}
\Return $G'$\;
\end{algorithm}

\subsection{Bytecode Compiler}
The computation graph is compiled into bytecode via two steps: shape tiling and bytecode encoding.

\begin{figure*}[t]
    \centering
    \includegraphics[width=0.9\linewidth]{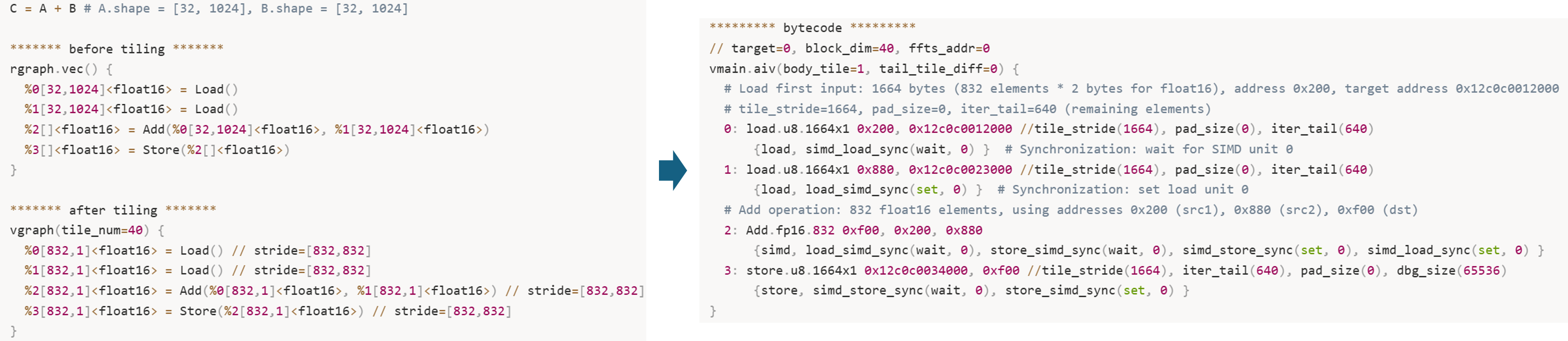}
    \caption{An example process of bytecode generation.}
    \label{fig:op compiling}
\end{figure*}

\textbf{Shape tiling.}
We partition the iteration space into multiple tiles to align with the hardware architecture.
The main idea of our tiling algorithm is that we design dedicated tiling templates for different types of (fused) operators and prune the tiling solutions with the hardware resource constraints and a lightweight cost model.
Specifically, as there are vector units and matrix units, we divide the computation operations into vector operations accordingly (e.g., $\mathsf{add}$) and matrix operations (e.g., $\mathsf{matmul}$). 

\Cref{alg: shape tiling} describes the detailed tiling algorithm for vector operations.
Given a computation graph $G$ consisting of vector operations, we first identify the dominant shape $S_d$ of $G$, i.e., the maximum number of dimensions and the maximum dimension sizes to cover the spatial dimensions of the operations in $G$ (line~1).
The tiling starts from the maximum tile size, $L$, i.e., the size of $S_d$, and then tries to decrease the tile size (line~11,15) by tiling $S_d$ from outer to inner (line~5,20).
The maximum tile size limit $T_{\mathsf{max}}$ is computed by dividing the local memory size by the peak number of live operators in $G$ (line~4).
Once we reach a dimension such that there can be possible tile sizes within the limit $T_{\mathsf{max}}$ (line~12), we select the best tile size from them using $\mathsf{hardware\_align\_div}$ based on a lightweight cost model considering the hardware alignment (line~14). %
Specifically, the cost of a tiling solution is measured by $\lceil \mathsf{\# tiles}/N \rceil* (\tilde{L} + 2)$, where $\mathsf{\# tiles}$ is the number of tiles, $\tilde{L}$ is the corresponding tile size, and 2 is used for the extra cost of computing a tile (e.g., decoding bytecode), so this formula models the bottleneck workload of all AI Cores.
After getting the minimum-cost tile size, $\mathsf{hardware\_align\_div}$ rounds it up to align with the hardware instruction width while respecting the $T_{\mathsf{max}}$ limit.
The remaining non-dominant operations are tiled accordingly (line~16-19).

{\color{black}For matrix operations, the output shape of each tile is 2-dimensional to align with the cube unit intrinsics; after tiling, we need to select the best swizzle strategy for the operations.
For the case where we fuse a matrix operation with multiple vector operations after it, we tile the iteration space of the matrix operation and the vector operations with the constraint that the vector operation can obtain its complete input data from the output of one matrix operation tile.}

\Cref{fig:op compiling} shows the example tiling solution for an element-wise $\mathsf{add}$ operation, which takes two float16 tensors of shape [32, 1024] as input. 
Suppose that this operator runs on an NPU with 40 AI Cores, and the hardware instruction width is 32 bytes.
For this operator, $\mathsf{hardware\_align\_div}$ first finds the minimum-cost tile size 820 (the tile number is 40 and the tail tile size is 788), and then rounds it up to 832 (the tail tile will not be padded).

\begin{figure}[t]
    \centering
    \includegraphics[width=0.9\linewidth]{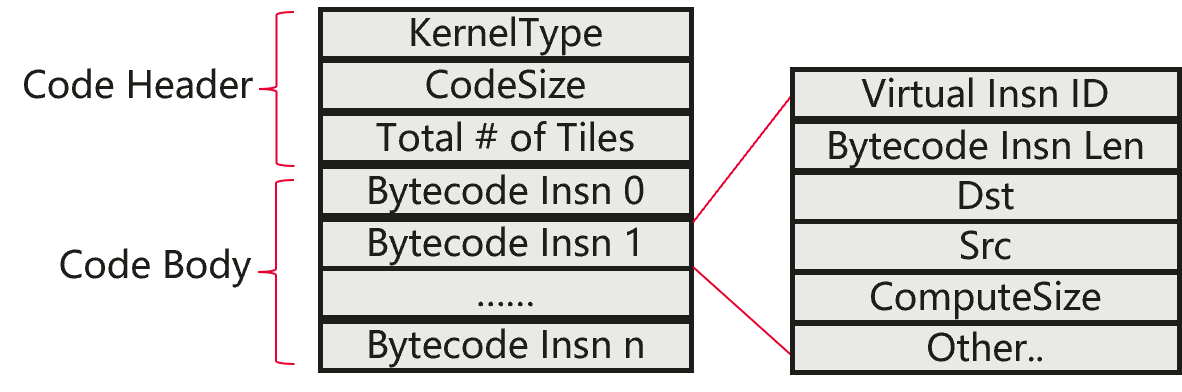}
    \caption{Bytecode Program Format.}
    \label{fig:bytecode program format}
\end{figure}

\textbf{Bytecode encoding.}
The tile computation will be encoded into a bytecode program.
For each fused operator, its bytecode program consists of two parts: the code header and the code body, as shown by~\Cref{fig:bytecode program format}. 
Specifically, the code body includes the bytecode of each operation in the fused operator.
Each operation bytecode encodes the necessary running information of the operation, including the operation type (the corresponding virtual instruction ID), the bytecode length (for later decoding), the input and output tile addresses (src and dst), the computation size ({\color{black}ComputeSize, corresponding to the ``size'' parameter for virtual instructions in~\Cref{tab:virtual_instruction}}), and other information, e.g., cond for Select in~\Cref{tab:virtual_instruction}. 
The code header includes the three kinds of information: 
(1) {\color{black}the kernel type (KernelType), e.g., a meta-kernel running on Vector units, or a stacking-based kernel that parallels multiple meta-kernels spatially (more details in~\Cref{sec: fusion})}; 
(2) the code size (CodeSize), i.e., the size of the {\color{black}code body}, used to determine the code body boundary and decode the operation bytecode,
and (3) {\color{black}the total number of tiles, used for AI Cores to determine their workloads}.
The bytecode program will be sent to the device for decoding and subsequent processing.

\Cref{fig:op compiling} prints an example bytecode program in readable format, from which we know the tile number (block\_dim), {\color{black}the number of tiles for each AI Core (body\_tile), and the kernel type (``vmain.aiv'' for Vector unit meta-kernels)}.
The program body contains the bytecode of each operation, together with the necessary synchronization operations that are automatically inserted.

\subsection{Virtual Machine}
The virtual machine is a kernel function on NPU that interprets bytecode by decoding it into the pre-defined virtual instructions and calling the corresponding virtual instruction functions.
Each AI Core runs a virtual machine.
The bytecode decoding is completed by the Scalar unit, and the virtual instruction functions are executed by the corresponding units.

\Cref{alg: vm} describes how the virtual machine works.
Given the bytecode program $P$ and the virtual instruction table $I$ that maps the virtual instruction ID to the corresponding virtual instruction function, the virtual machine first computes the number of tiles assigned to each AI Core (line~4) and then determines the tile range for it (line~5), based on its ID $\mathsf{id}$, the total number of AI Cores $N$ to use, and the total number of tiles $M$ specified by $P$.
For each tile in the range, the virtual machine interprets all the bytecodes in the code body of $P$.
Specifically, a bytecode $b$ is extracted based on its starting address $p$ and its length $\mathsf{Insn\_Len}$.
We start from the address of the code body of $P$ to process the first bytecode and stop when reaching the code body boundary (line~6,7, 14).
Each bytecode corresponds to a virtual instruction function call $f()$ to finish the workload (line~9-13). If the bytecode is about memory operations, $f$ also requires the tile ID as its input to load/store data to the correct global memory address (line~11).

In practice, \ours~can hide the decoding overhead well, because as shown by~\Cref{fig:comp flow}, (1) the bytecode decoding and the instruction execution can be pipelined, and (2) the decoding, which runs scalar computation, is much faster than the vector/matrix/{\color{black}DMA instructions}.

\begin{algorithm}[t]
\caption{Virtual Machine}
\label{alg: vm}
\KwIn{Bytecode program $P$, virtual instruction table $\mathcal{I}$}
$\mathsf{id} \leftarrow$ the ID of the current AI Core\;
$N \leftarrow$ the total number of AI Cores to run $P$\;
$M \leftarrow$ the total number of tiles specified by $P$\;
$m \leftarrow \lceil M/N \rceil$\;
\For{$i \in [ m*\mathsf{id}, \min{(M, m*(\mathsf{id}+1)} )$}{
    $p,p_0\leftarrow$ the address of the code body of $P$\;
    \While{$p<p_0+P.\mathsf{CodeSize}$}{
        $b\leftarrow$ the operation bytecode at $p$\;
        $f \leftarrow I[b.\mathsf{Insn\_ID}]$; {\small \color{olive}// Virtual Insn function of $b$}
        
        \If{$f$ is memory operation}{
            $f(b, i)$; {\small \color{olive}// memory operation requires tile ID}
        }
        \Else{
            $f(b)$\;
        }
        $p\leftarrow p+b.\mathsf{Insn\_Len}$\;
    }
}

\end{algorithm}

\section{Operator Fusion}~\label{sec: fusion}

In this section, we first introduce the fusion categories that we support to increase the fusion opportunities.
Then, we introduce how to fuse operators on different types of computation graphs.

\subsection{Fusion Categories}

The runtime operator fuser supports two types of operator fusion, i.e., pattern-based fusion and stacking-based fusion, which are flexible and general enough to provide a lot of fusion opportunities.

\begin{figure}[t]
    \centering
    \includegraphics[width=0.4\linewidth]{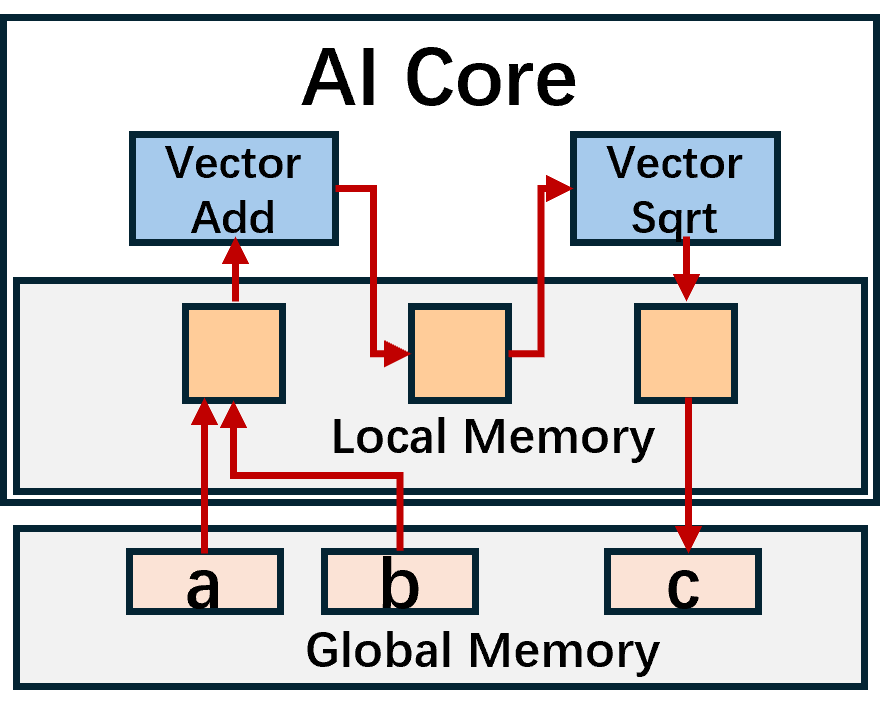}
    \caption{Fusing two vector operations via local memory.}
    \label{fig:fuse vector}
\end{figure}

\textbf{Pattern-based fusion} tries to merge the iterations of multiple operators based on specific patterns. %
There are two typical fusion patterns for the Ascend NPU: (1) fusion between vector operations, (2) fusion between cube and vector operations.
\Cref{fig:fuse vector} shows an example of vector-vector fusion. 
Specifically, the original operators are $c = a+b, c = \sqrt{c}$. By fusing the two operators, the result of $a+b$ is stored in the local memory instead of the global memory for the following $\mathsf{sqrt}$ operation to read, therefore reducing the expensive global memory access.
For the cube-vector fusion, we currently only support fusing a cube operation with multiple element-wise vector operations after it. The Vector and the Cube units can run in parallel in this case to overlap the computation latency.
The kernels generated through pattern-based fusion, including the kernel of a single unfused operator, are called meta-kernels.

\begin{figure}[t]
    \centering
    \includegraphics[width=0.6\linewidth]{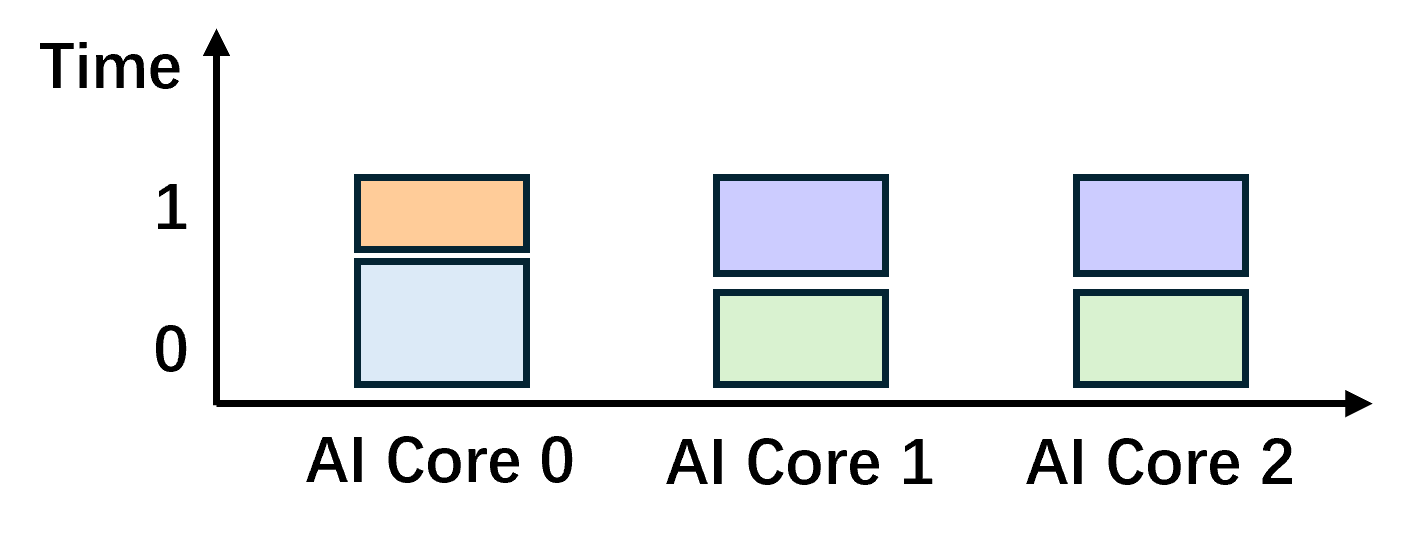}
    \caption{Stacking kernels spatially and temporally.}
    \label{fig:stacking}
\end{figure}

\textbf{Stacking-based fusion} {\color{black}stacks different meta-kernels} temporally and spatially to reduce latency (\Cref{fig:stacking}).
The spatial stacking schedules the tile computation of independent operators to different AI Cores for higher computation resource utilization.
The temporal stacking schedules different operator tiles to the same AI Core for sequential execution, to reduce the runtime scheduling overhead.
The temporal stacking does not require operators to be independent.
In practice, we can combine the two types of stacking for higher efficiency.

\subsection{Fusion on Static/Dynamic Graphs}
We now introduce how we fuse operators on static and dynamic computation graphs, respectively.

\textbf{Static computation graphs.} 
The graph structure of the static computation graph is fixed and known before execution. 
Therefore, despite the lack of concrete operator shape information, we can {\color{black}check the fusion conditions, e.g., the iterations of two operators are mergeable, based on symbol deduction} to make fusion decisions.
Given a computation graph, we first group the basic operators into clusters to reduce the fusion complexity.
For the subgraph of each basic operator cluster, we split it into multiple fused subgraphs based on pre-defined fusion rules.
Specifically, for a cube operation, we will try to fuse it with the element-wise vector operations after it (if any).
For vector operations, if their iteration spaces can be merged, we can fuse them, which leads to a pattern-based fusion; otherwise, {\color{black}we can apply stacking-based fusion to them}.

\textbf{Dynamic computation graphs.} 
For dynamic computation graphs whose graph structure is uncertain before execution, we make fusion decisions in a streaming way.
Specifically, when a dynamic model is running, the operators that can be determined will be added to a buffer, and the fusion decisions will be made immediately {\color{black}based on the same rules for static graphs}. 
The fused operators will be flushed to the operator compiler under some conditions, e.g., the newly added operator cannot be fused, the computation result has to be sent back to the Host side (like printing the result).

\begin{figure}[t]
    \centering
    \includegraphics[width=0.5\linewidth]{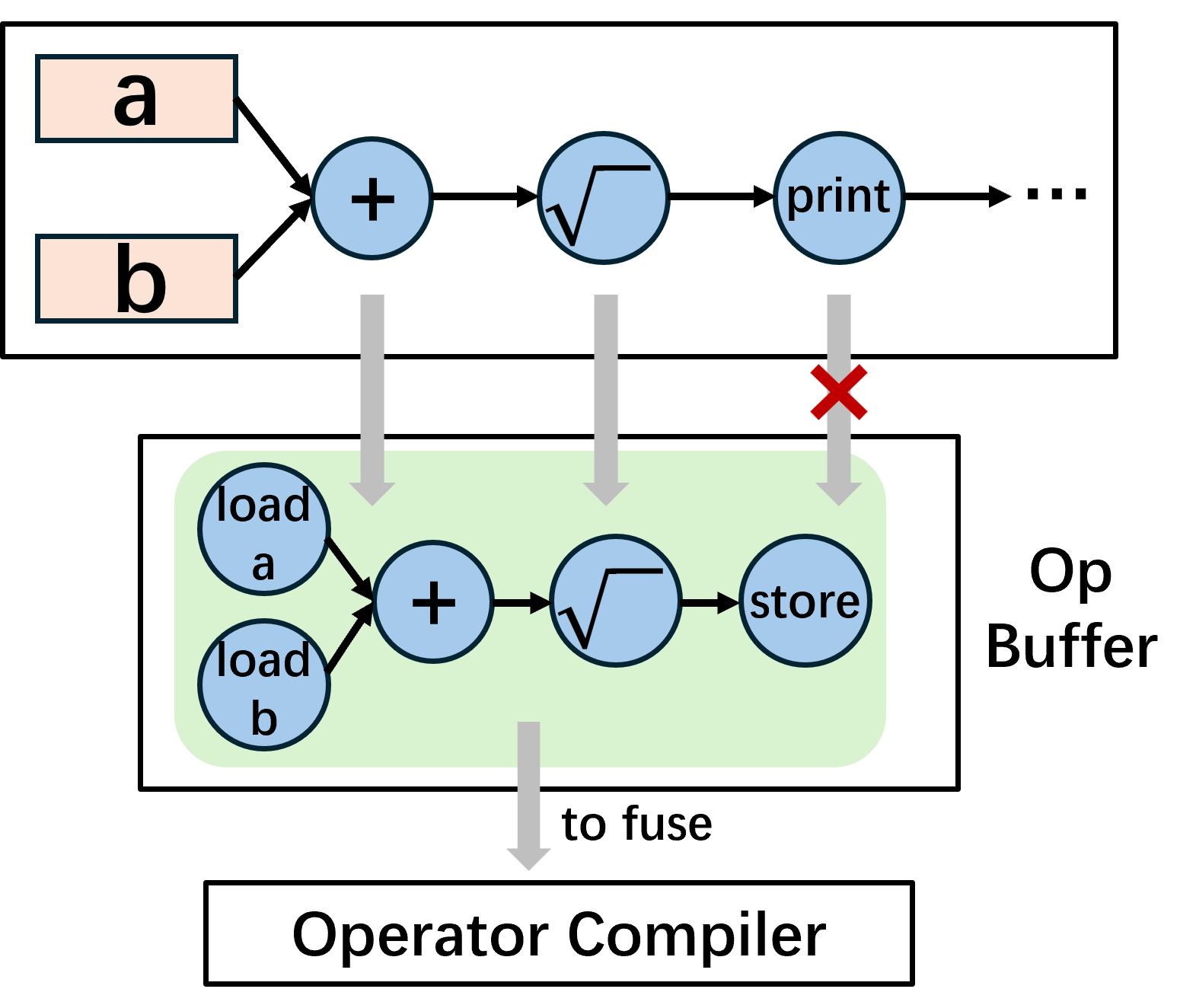}
    \caption{{\color{black}An example of operator fusion on dynamic graphs.}}
    \label{fig:op fuser example}
\end{figure}

\Cref{fig:op fuser example} shows a running example of the operator fuser on dynamic graphs.
In the example, a part of the model is to first compute $c = a+b$, $d = \sqrt{c}$, and then print the value of $d$, where all the variables are tensors.
At runtime, the $\mathsf{add}$ operator and the $\mathsf{sqrt}$ operator are added to the buffer in order, {\color{black}with the necessary $\mathsf{load}$ operators being added as well}.
We can fuse the $\mathsf{add}$ and the $\mathsf{sqrt}$ operators according to the vector-vector fusion pattern.
However, the $\mathsf{print}$ operator requires the square root result to be sent back to the Host, so the fuser stops fusing, {\color{black}adds the necessary $\mathsf{store}$ operator,} and flushes the fused operator to the operator compiler for compiling and execution.

\section{Experiment}

In this section, we want to answer three questions:
(1) Can \ours~generate efficient implementations for (fused) operators?
(2) Can the operator fuser fuse operators effectively?
(3) Is the compilation overhead light enough to make the end-to-end running of models efficient?

\textbf{Datasets:}
We evaluate \ours~on the operator, the subgraph, and the model level, respectively.
For operator-level analysis, we focus on the dynamic $\mathsf{matmul}$ operator. 
For subgraph-level analysis, we test three compound operators: $\mathsf{LayerNorm}$, $\mathsf{addmm}$, and $\mathsf{if}$-$\mathsf{else}$-$\mathsf{add}$ ($(a>b~?~2x:4x)+y$), where $\mathsf{if}$-$\mathsf{else}$-$\mathsf{add}$ has both a dynamic shape and a dynamic control flow.
For model-level analysis, we test two small models, BERT (in training mode) and MMoE (in inference mode), as well as two LLMs, Qwen3-14B and Llama3.1-8B (both in finetuning mode).
The details of the dynamic shape ranges of the 8 test datasets are listed in~\Cref{tab:shapes_and_ranges}.
We randomly sample 60 shapes for each dynamic operator/subgraph. For $\mathsf{if}$-$\mathsf{else}$-$\mathsf{add}$, we also randomly set the condition value to make it execute different branches.
For the dynamic models, we enumerate all the shapes in the possible shape sets.

Qwen3-14B runs with the model parallelism degree of 8 on 8 NPUs, and Llama3.1-8B runs with the data parallelism degree of 2 and the model parallelism degree of 4 on 8 NPUs, while other experiments are performed on a single NPU.

\begin{table*}[t]
  \centering
  \caption{Dynamic shape ranges ($\mathcal{D}$ are shapes extracted from common LLMs)}
  \label{tab:shapes_and_ranges}
  \vspace{6pt}
  \resizebox{0.8\textwidth}{!}{%
  \begin{tabular}{l l l}
    \toprule
    \textbf{Name} & \textbf{Input Shape} & \textbf{Shape Range} \\
    \midrule
    matmul  & $([m,k], [k,n])$ & $m \in [1, 8192]$, $(n,k) \in \mathcal{D} =\{(4096, 4096), (11008, 4096), (4096, 11008),$ \\
           & & $(5120, 5120), (13696, 5120), (5120, 13696), (8192, 8192),$ \\
           & & $(28672, 8192), (8192, 28672)\}$ \\
    \midrule
    LayerNorm & $([b, s, h])$ & $b \in [1, 60]$, $s = 8192$, $h \in \{1024, 2048, 3072, 4096\}$ \\
    \addlinespace
    addmm & $([m,k], [k,n], [m,n])$ & $m \in [1, 8192]$, $(n,k) \in \mathcal{D}$ \\
    \addlinespace
    if-else-add & $([b, s, f])$ & $b \in [1, 256]$, $s \in [1, 512]$, $f \in [1, 8192]$ \\
    \midrule
    BERT (Train) & $([b, s])$ & $b \in \{2, 4, 8, 16\}$, $s = 128$ \\
    \addlinespace
    MMoE (Inference) & $([b,f])$ & $b \in \{2048, 4096, 8192, 16384\}$, $f=47104$ \\
    \addlinespace
    Qwen3-14B (Finetune) & $([b,s])$ & $b \in \{1, 2, 4, 8\}$, $s = 4096$ \\
    \addlinespace
    Llama3.1-8B (Finetune) & $([b,s])$ & $b \in \{1, 2, 3, 4\}$, $s = 8192$ \\
    \bottomrule
  \end{tabular}
  }
\end{table*}

\textbf{Baselines:}
Because most of the existing works are not targeted and implemented for Ascend NPUs, we only include the available Ascend NPU compilers as the baselines for comparison.
Specifically, we have 4 baselines in our experiments:

{\color{black}\noindent(1) \textbf{\eagerfull~(\eager)}: eager mode of torch-npu~\cite{torchnpu} (the version of PyTorch 2~\cite{ansel2024pytorch} running on Ascend NPUs), based on AOL kernels~\cite{aol} without automatic operator fusion.

\noindent(2) \textbf{\tritonrfull~(\tritonr)}: the recompilation mode of the \tritonfull~compiler in torch-npu~\cite{torchnpu} adapted from PyTorch 2~\cite{ansel2024pytorch}, which accepts the complete computation graph of a dynamic model to determine operator fusion before execution and recompiles the operator for each new shape encountered.

\noindent(3) \textbf{\tritondfull~(\tritond)}: the dynamic mode of the \tritonfull~compiler in torch-npu~\cite{torchnpu} adapted from PyTorch 2~\cite{ansel2024pytorch}. It performs the same operator fusion as \tritonr, but only compiles a dynamic operator once, based on the size assumption it generates on the first input of the operator. 
When the size assumption does not hold, a just-in-time compilation will be triggered.

\noindent(4) \textbf{\msfull~(\ms)}: MindSpore graph O0 mode~\cite{mindspore-graph}, based on AOL~\cite{aol}, no automatic operator fusion.}

When compiling operators, both \tritonr~and \tritond~will generate multiple candidate kernels and benchmark them on the NPU.
Both \eager~and \ms~do not have compilation overhead, as they do not perform operator fusion and operator compilation.

For comparison, we implement \ours~as a compiler backend of torch-npu~\cite{torchnpu} and MindSpore~\cite{ms}, {\color{black}denoted by \textbf{\ourspt} and \textbf{\oursms} respectively}.
All the operator and subgraph experiments are conducted on torch-npu, while the model experiments are run on both torch-npu and MindSpore.
{\color{black}Based on the implementation in the current stage}, we evaluate the small models on torch-npu and the LLMs on MindSpore.
For the model-level evaluation on torch-npu, since the operator fuser of \ours~has not been integrated into torch-npu currently, \ourspt~follows the operator fusion decisions as \tritonr~and \tritond; for the LLM evaluation on MindSpore, \oursms~uses our static graph operator fuser.

\textbf{Metrics:}
We use two metrics in the evaluation, i.e., (1) the \textbf{\textit{running time}} of an operator/subgraph/model instance {\color{black}(the host side cost in running is counted, e.g., computing buffer shapes dynamically)} and (2) the corresponding \textbf{\textit{compilation time}}.
For \tritonr~and \tritond, the running time does not contain the compilation time because compilation is done before running a test case.
For \ours, the compilation process is mixed with the running process. 
For analysis, we only collect the host-side time cost as the compilation time of \ours.
The running time of an operator/subgraph is the average latency of 100 executions, while for models, it is the average latency of 10 executions.
To eliminate artifacts from consecutive runs, the L2 cache is cleared before each execution.

\textbf{Testbed:}
We mainly obtain the evaluation results on an openEuler 22.03 (LTS-SP4) machine with 4 64-core Kunpeng-920 CPUs (256 logical processors in total), 8 Ascend 910B2 NPUs, and 2TB RAM.
The results of BERT, MMoE, and Qwen3-14B are obtained on an openEuler 22.03 (LTS-SP4) machine with 4 48-core CPUs (384 logical processors in total), 8 Ascend 910 NPUs, and 2TB RAM, where the 910 NPUs have a similar architecture to that of 910B NPUs but with higher computation power.
The experiments are run with torch-npu 2.7.1, CANN 8.5.0, triton-ascend 3.4.0, and MindSpore 2.7.2 (for Llama3.1-8B) / MindSpore 2.8.0 (for Qwen3-14B).

\begin{figure}[t]
    \centering
    \includegraphics[width=1\linewidth]{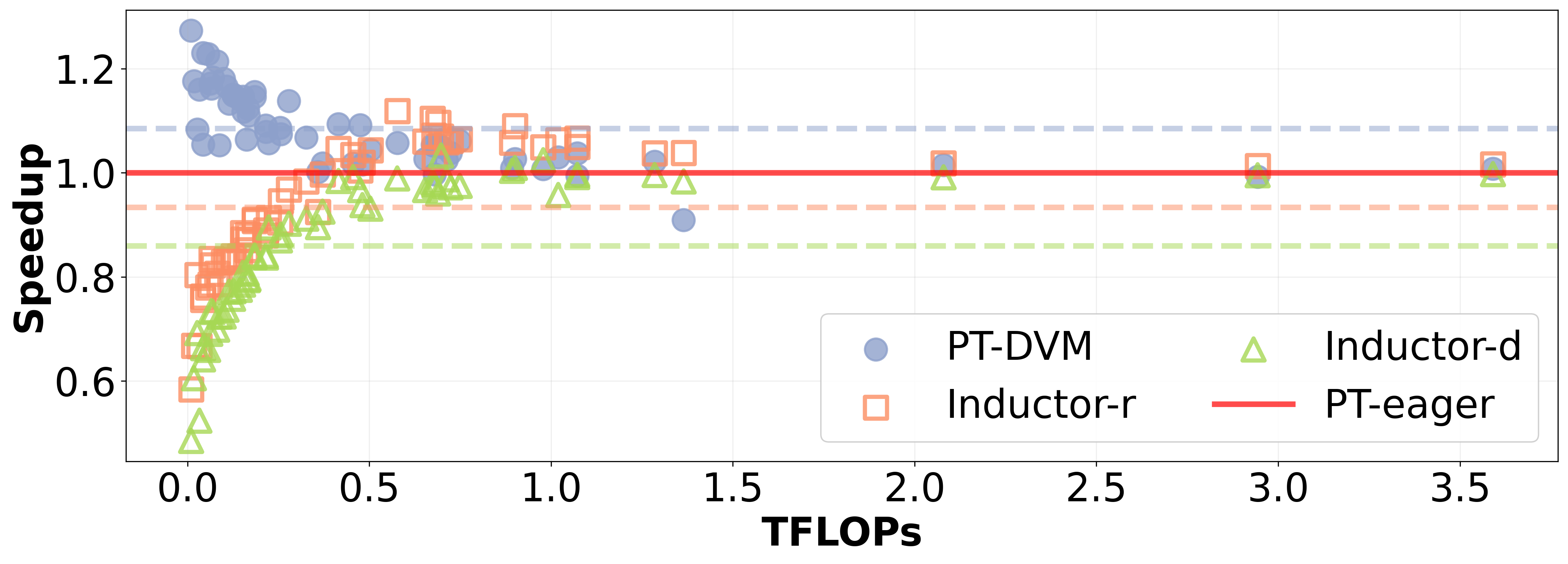}
    \caption{matmul}
    \label{fig:matmul}
\end{figure}

\begin{figure}[t]
    \centering
    \includegraphics[width=1\linewidth]{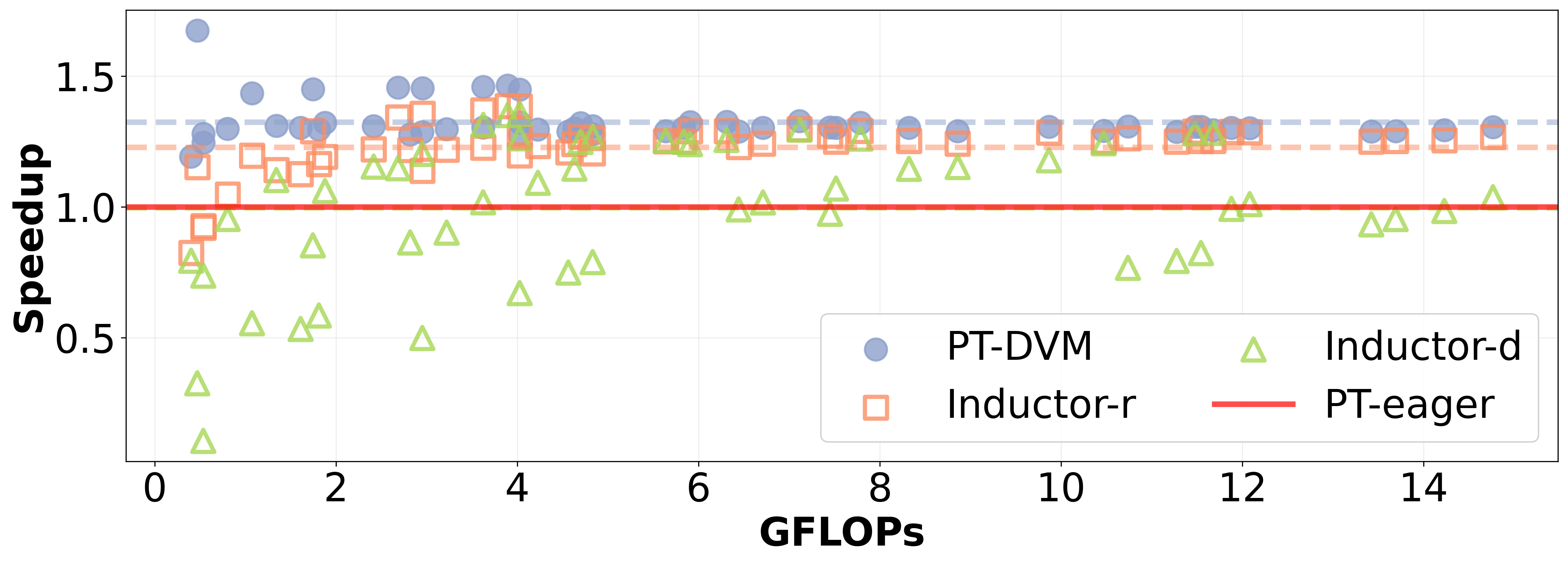}
    \caption{LayerNorm}
    \label{fig:layernorm}
\end{figure}

\begin{figure}[t]
    \centering
    \begin{subfigure}[b]{1\linewidth}
        \centering
        \includegraphics[width=\linewidth]{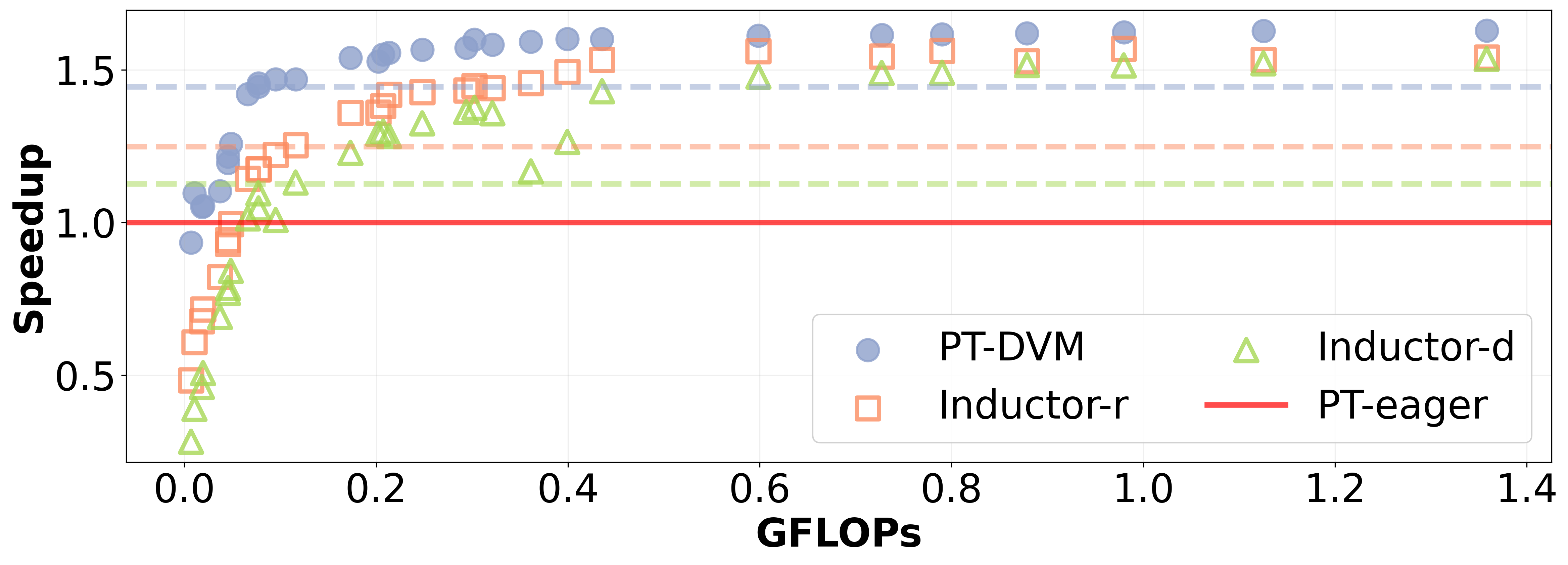}
        \caption{``True'' branch}
        \label{fig:ifelse true}
    \end{subfigure}
    \hfill
    \begin{subfigure}[b]{1\linewidth}
        \centering
        \includegraphics[width=\linewidth]{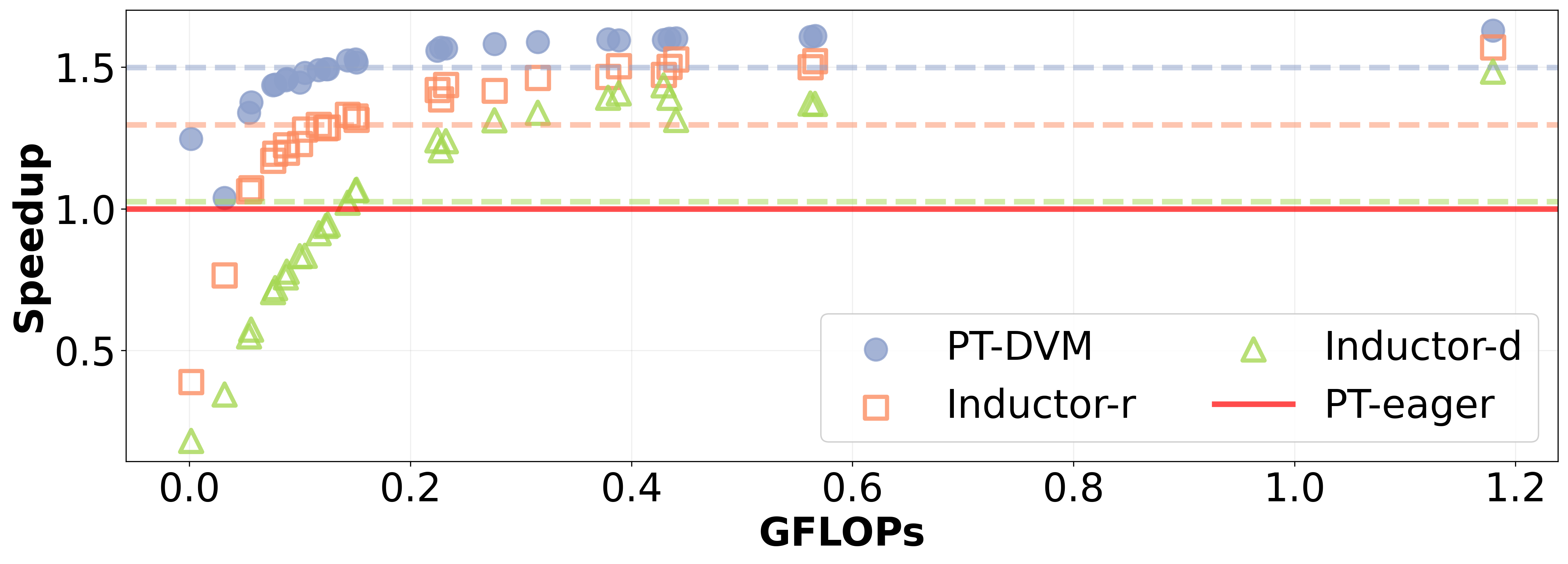}
        \caption{``False'' branch}
        \label{fig:ifelse false}
    \end{subfigure}
    \caption{if-else-add}
    \label{fig:ifelse}
\end{figure}

\begin{figure}[t]
    \centering
    \includegraphics[width=1\linewidth]{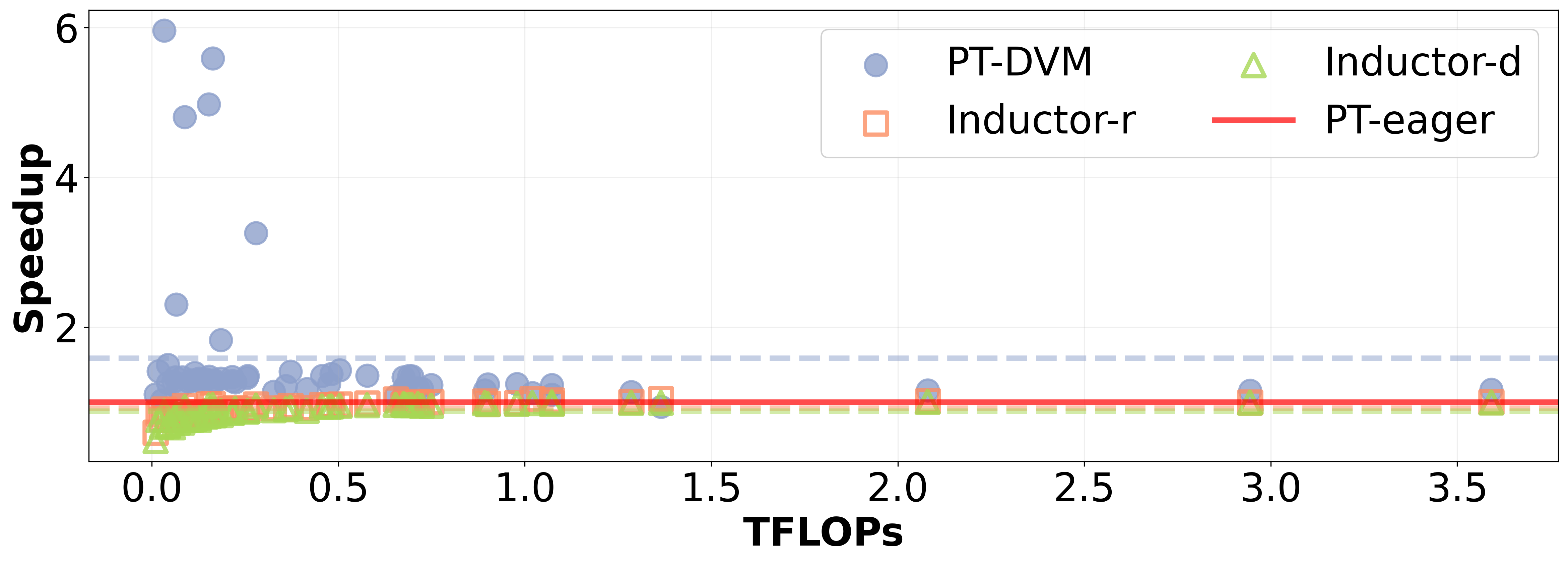}
    \caption{addmm}
    \label{fig:addmm}
\end{figure}

\begin{figure}[t]
    \centering
    \includegraphics[width=1\linewidth]{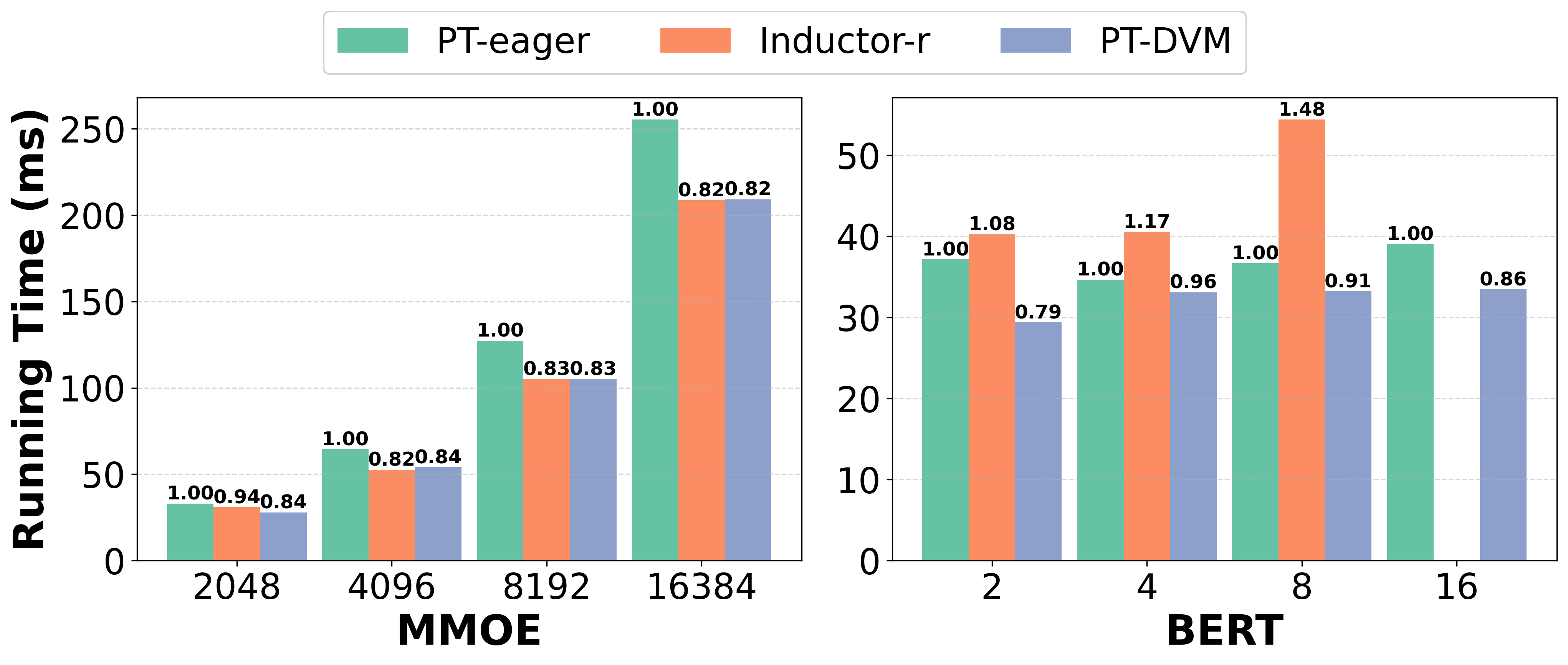}
    \caption{ML models}
    \label{fig:ml models}
\end{figure}

\begin{figure}[t]
    \centering
    \includegraphics[width=1\linewidth]{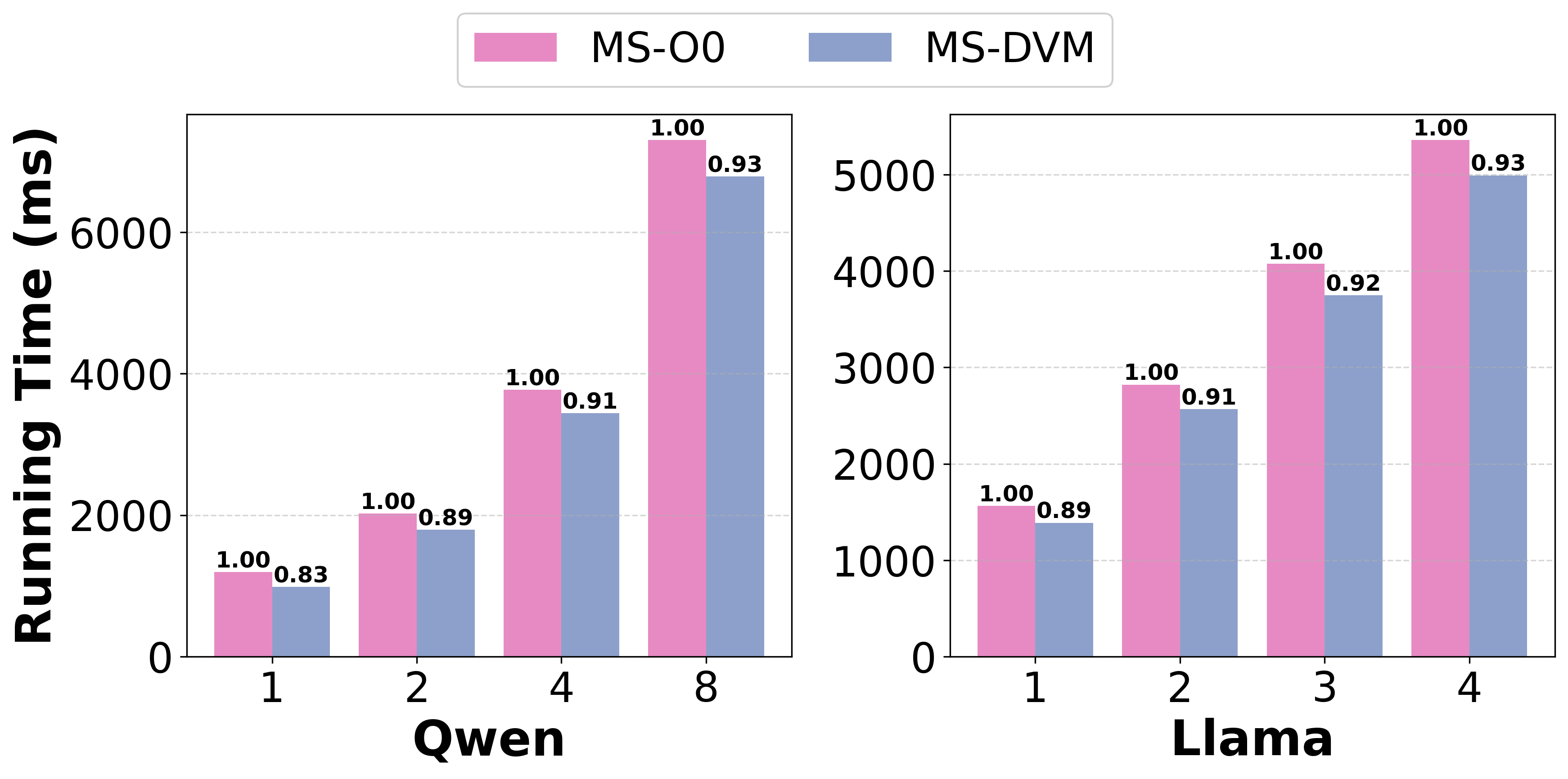}
    \caption{LLM}
    \label{fig:llms}
\end{figure}

\subsection{Running Time Comparison}

\subsubsection{Single Operator}

\Cref{fig:matmul} shows the speedups of all the methods, i.e., \ourspt, \tritonr, and \tritond, against \eager~on $\mathsf{matmul}$ in terms of the operator running time, where each point corresponds to a sampled operator instance, and the dashed lines show the average speedups of the corresponding methods.
\Cref{tab:speedup} shows the running time speedups of \ourspt~against the baselines.
{\color{black}For $\mathsf{matmul}$, all the baselines run the corresponding AOL kernel~\cite{aol}. However, the results of these baselines in~\Cref{tab:speedup} are different, due to the host side cost difference.
For example, \tritonr~treats the operator instances as static, while \tritond~not, resulting in extra running time for operations such as computing the output buffer shape for each operator instance; on the other hand, \tritonr~and \tritond~perform more operations than \eager~to launch a kernel, incurring higher host side cost.}
Compared with the three baselines, \ourspt~is 0.88-2.62$\times$ better, with an average speedup of 1.09-1.31$\times$.
The results demonstrate the ability of \ours~to find high-performance operator implementation efficiently, because the time of bytecode generation and decoding is included in the running time of \ourspt.

\begin{table}[t]
\centering
\caption{\ourspt~Running Time Speedups against the baselines.}
\label{tab:speedup}
\resizebox{\linewidth}{!}{%
\begin{tabular}{l l S[table-format=1.2] c c}
\toprule
\textbf{Name} & \textbf{Method} & \textbf{Avg. Speedup} & \textbf{Speedup Range} & \multicolumn{1}{c}{\textbf{Ratio of}} \\
 & & & & \multicolumn{1}{c}{\textbf{speedup (\textgreater 1)\%}} \\
\midrule
\multirow{3}{*}{matmul} 
& \tritonr & 1.19 & 0.88 - 2.18 & 62 \\
& \tritond & 1.31 & 0.93 - 2.62 & 93 \\
& \eager & 1.09 & 0.91 - 1.27 & 93 \\
\midrule
\multirow{3}{*}{LayerNorm} 
& \tritonr & 1.09 & 1.01 - 1.45 & 100 \\
& \tritond & 1.63 & 1.01 - 11.77 & 100 \\
& \eager & 1.32 & 1.19 - 1.67 & 100 \\
\midrule
\multirow{3}{*}{if-else-add} 
& \tritonr & 1.21 & 1.03 - 3.21 & 100 \\
& \tritond & 1.58 & 1.06 - 6.88 & 100 \\
& \eager & 1.47 & 0.93 - 1.63 & 98 \\
\midrule
\multirow{3}{*}{addmm} 
& \tritonr & 1.73 & 0.90 - 6.59 & 98 \\
& \tritond & 1.82 & 0.94 - 6.81 & 98 \\
& \eager & 1.59 & 0.94 - 5.96 & 98 \\
\bottomrule
\end{tabular}
}
\end{table}

\begin{table}[t]
\centering
\caption{Op \& subgraph compilation time comparison (time in ms). CT: compilation time; RT: running time.} 
\label{tab:compilation time ops}
\resizebox{\linewidth}{!}{
\begin{tabular}{@{} l l S[table-format=6.2] S[table-format=7.2] S[table-format=5.2] S[table-format=2.2e2] @{}}
\toprule
\textbf{Name} & \textbf{Method} & \textbf{Max CT} & \textbf{Total CT} & \textbf{Total RT} & \multicolumn{1}{c}{\textbf{Total CT /}} \\
 & & & & & \multicolumn{1}{c}{\textbf{Total RT}} \\
\midrule
\multirow{4}{*}{matmul} 
    & \ourspt      & 0.11   & 0.48    & 123.26 & 3.89e-3 \\
    & \tritonr & 115.88 & 4339.33 & 125.92 & 3.45e1 \\
    & \tritond & 200.99 & 354.53  & 133.79 & 2.65e0 \\
\midrule
\multirow{4}{*}{LayerNorm} 
    & \ourspt      & 0.10    & 1.51       & 271.14 & 5.57e-3 \\
    & \tritonr & 37802.36 & 1517938.56 & 282.74 & 5.37e3 \\
    & \tritond & 35696.20 & 108446.42  & 357.43 & 3.03e2 \\
\midrule
\multirow{4}{*}{if-else-add} 
    & \ourspt      & 0.05    & 2.12       & 97.79 & 2.17e-2 \\
    & \tritonr & 35352.37 & 1683099.71 & 106.74 & 1.58e4 \\ 
    & \tritond & 35425.44 & 66679.14   & 120.13 & 5.55e2 \\
\midrule
\multirow{4}{*}{addmm} 
    & \ourspt      & 0.11   & 1.41    & 127.12 & 1.11e-2 \\
    & \tritonr & 521.79 & 6242.67 & 168.29 & 3.71e1 \\
    & \tritond & 288.30 & 444.56  & 172.67 & 2.58e0 \\
\bottomrule
\end{tabular}
}
\end{table}

\subsubsection{Subgraph}
\Cref{fig:addmm,fig:layernorm,fig:ifelse} show the running time speedups of different methods against \eager.
In terms of operator fusion, \ours~can successfully fuse the operators for all the tested subgraphs; \tritonr~and \tritond~can fuse the operators for $\mathsf{LayerNorm}$ and $\mathsf{if}$-$\mathsf{else}$-$\mathsf{add}$, however, for $\mathsf{addmm}$ they directly call the vendor-provided AOL kernels~\cite{aol}, which do not fuse the operators in it;
\eager~calls the AOL kernels for all the subgraphs, which only fuse the operators in~$\mathsf{LayerNorm}$, leaving other subgraphs unfused.

\Cref{tab:speedup} shows that, for $\mathsf{addmm}$ operators, \ourspt~beats \eager, \tritonr, and \tritond~on 98\% of the sampled instances.
Thanks to operator fusion, the maximum speedups of \ourspt~against the baselines are 5.96-6.81$\times$, and the respective average speedups are 1.59-1.82$\times$.
\tritonr, \tritond~and \eager~have similar performance because they all call the AOL kernels.
On $\mathsf{LayerNorm}$, although all the methods run a fused operator, their performances are different. 
Specifically, \ourspt~is consistently better than all the baselines and \tritonr~outperforms \eager~in most cases.
However, \tritond~can only beat \eager~on 56\% of the sampled instances, showing that the generated kernel of \tritond~does not fully cover the possible shape range for $\mathsf{LayerNorm}$.
On $\mathsf{if}$-$\mathsf{else}$-$\mathsf{add}$,~\Cref{fig:ifelse} reports the speedups of different methods when the ``if'' condition holds or not separately.
All the methods except \eager~can fuse the multiplication and the $\mathsf{add}$ for each branch (\tritonr~and \tritond~will try to recompile the subgraph when new execution paths are activated), therefore, they all outperform \eager~when the FLOPs is large.
For \ourspt, since it needs to generate the correct $\mathsf{if}$-$\mathsf{else}$-$\mathsf{add}$ function definition each time a sampled instance is tested, its compilation time and hence the running time is increased, making it perform slightly worse than \eager~when the computation FLOPs is small (less than 0.007ms).
\tritond~performs worse than \ourspt~and \tritonr~due to the lack of concrete shapes.
Compared with \tritonr, \tritond~and \eager, \ourspt~is 1.21$\times$, 1.58$\times$ and 1.47$\times$ better on average.

\subsubsection{Model}
\Cref{fig:ml models,fig:llms} compare the model running time of different methods, i.e., the time to run a batch of requests.
\eager~and \ms~both call AOL kernels and do not perform graph-level optimizations. 
\tritonr~treats the dynamic models as static models with all shape information known.
When the batch size is 16 for BERT, \tritonr~encounters an OOM problem in the NPU local memory, so we do not report its results on this case.
The figures show that \ours~consistently outperforms or achieves close performance compared with the other methods on all the model instances.
Specifically, \ourspt~is 1.05-1.26$\times$ better than \eager, and 0.97-1.64$\times$ better than \tritonr.
Since \ourspt~share the same graph-level optimizations as \tritonr~in the model-level evaluation, this result demonstrates the effectiveness and efficiency of the operator optimization of \ours.
On BERT, as the batch size increases, there is a clear increase in the running time for \tritonr, while the running time of both \eager~and \ourspt~remain relatively stable, which indicates that for \eager~and \ourspt, the NPU hardware resources are well utilized and have not been saturated by the computation workloads.
Compared with \ms, \oursms~is 1.07-1.21$\times$ better, demonstrating the graph-level optimization capability of \ours~on static graphs as well as the dynamic operator optimization capability of \ours.

\subsection{Compilation Time Comparison}

\Cref{tab:compilation time ops,tab:compile_times_ml,tab:compile_time_llm} compare the compilation time of different methods in detail.
Overall, the compilation time of \ours~is negligible; by contrast, \tritonr~and \tritond~both perform multiple hardware candidate kernel measurements, {\color{black}heavy graph-level analysis and a long compilation flow (e.g., many dialect conversions)}, which is time-consuming. 

Specifically, on $\mathsf{matmul}$, the maximum compilation time of \ourspt~for an operator instance is 0.11 ms, and the ratio of the total compilation time to the total running time is 0.389\%, while the respective maximum compilation times of \tritonr~and \tritond~are 115.88 ms and 200.99 ms for $\mathsf{matmul}$.
Note that for $\mathsf{matmul}$, \tritonr~and \tritond~choose to directly use the vendor-provided library with no operator compilation, so the compilation time for these two methods is simply the time to go through the complete compilation process.
Since \tritonr~compiles each operator instance it encounters, while \tritond~generates one kernel for different operator shapes, the total compilation time of \tritonr~is longer than that of \tritond. 
The total compilation times of \tritonr~and \tritond~are 34.5$\times$ and 2.65$\times$ of the total running time for $\mathsf{matmul}$ for the 60 sampled operator instances, respectively.
Such compilation time can be unacceptable when there are multiple possible shapes or when there is a requirement of quick deployment of models on new hardware (even if \tritond~only compiles a dynamic operator once, this total compilation time will be long if there are many dynamic operators to optimize).

For the subgraph-level evaluation, the compilation time of \tritonr~and \tritond~increases to tens of seconds on $\mathsf{LayerNorm}$ and $\mathsf{if}$-$\mathsf{else}$-$\mathsf{add}$, {\color{black}because for these two kinds of operators, the operator compilation is triggered}.
The respective compilation time ratios of \tritonr~and \tritond~to the total running time are up to {\color{black}1.58e4 and 5.55e2}.
Note that \tritond~generates two kernels for $\mathsf{if}$-$\mathsf{else}$-$\mathsf{add}$, one per branch, while for $\mathsf{LayerNorm}$, \tritond~assumes the reduction dimension is static, therefore the just-in-time compilation is triggered four times in total, one for each reduction dimension size (i.e., the value of $h$ in the input shape in~\Cref{tab:shapes_and_ranges}). 
By contrast, \ourspt~remains efficient on all the test cases, spending less than 2.17\% of the total running time in compilation.
The ratio 2.17\% appears on $\mathsf{if}$-$\mathsf{else}$-$\mathsf{add}$, because generating the correct function definition with the selected branch will increase the overhead. However, this ratio is still negligible. %

For the model-level evaluation, the compilation time of \ourspt~is the cost of the graph-level compilation in \tritonr, which is much longer compared with the operator compilation time of \ourspt~(we can infer this by comparing the running time, where the operator optimization time is included, with the compilation time).
The compilation time of \tritonr~includes both the graph- and the operator-level optimization time, which is up to 3760s.
By comparing \tritonr~ and \ourspt, the efficiency of our operator compiler can be validated again.
On MindSpore, the compilation time of \oursms~is the graph-level compilation time of it. The result that \oursms~and \ms~have close compilation time means that the static graph compilation by \ours~is efficient as well.
Note that although the ratio of the compilation time to the model running time is up to $1.28\times10^3$ for \ours~(on BERT), the compilation time, i.e., the graph-level optimization time, can be amortized given that (1) the reported model running time is per batch but a model can be run for multiple times in practice, and (2) the graph-level optimization result can be reused since these models are static graphs.

\begin{table}[t]
\centering
\caption{Compilation time comparison for MMoE and BERT.}
\label{tab:compile_times_ml}
\resizebox{\linewidth}{!}{%
\begin{tabular}{@{} l l c c @{}}
\toprule
Group & Method & MMoE & BERT \\
\midrule
\multirow{1}{*}{\shortstack{Compilation \\ Time (ms)}} & \tritonr & $4.84\times10^5$ - $3.76\times10^6$ & $1.50\times10^5$ - $3.43\times10^5$ \\
 & \ourspt & $2.71\times10^1$ - $2.04\times10^2$ & $4.09\times10^4$ - $4.23\times10^4$ \\
\midrule
\multirow{1}{*}{\shortstack{Compilation/ \\ Running Time}} & \tritonr & $1.28\times10^4$ - $1.70\times10^4$ & $3.00\times10^3$ - $9.24\times10^3$ \\
 & \ourspt & $7.97\times10^{-1}$ - $8.20\times10^{-1}$ & $1.11\times10^3$ - $1.28\times10^3$ \\
\bottomrule
\end{tabular}
}
\end{table}

\begin{table}[t]
\centering
\caption{Compilation time comparison for Qwen and Llama.}
\label{tab:compile_time_llm}
\resizebox{\linewidth}{!}{%
\begin{tabular}{@{} l l c c @{}}
\toprule
Group & Method & Qwen & Llama \\
\midrule
\multirow{1}{*}{\shortstack{Compilation \\ Time (ms)}} & \ms & $2.66\times10^2$ - $8.41\times10^2$ & $1.18\times10^2$ - $1.29\times10^2$ \\
 & \oursms & $2.78\times10^2$ - $3.02\times10^2$ & $1.27\times10^2$, $1.28\times10^2$ \\
\midrule
\multirow{1}{*}{\shortstack{Compilation/ \\ Running Time}} & \ms & $1.15\times10^{-1}$ - $2.23\times10^{-1}$ & $2.41\times10^{-2}$ - $7.55\times10^{-2}$ \\
 & \oursms & $2.81\times10^{-1}$ - $4.45\times10^{-2}$ & $2.55\times10^{-2}$ - $9.20\times10^{-2}$ \\
\bottomrule
\end{tabular}
}
\end{table}

\section{Related Work}

\textbf{Dynamic Shape Operator optimization.}
Existing works on optimizing dynamic shape operators can be divided into 2 categories: 
(1) One-kernel-per-shape~\cite{xla,suhan2021lazytensor,torch-xla} and (2) multi-kernels~\cite{tvm_bucketing,zheng2022dietcode,zheng2023bladedisc,zhou2025sample,cublas,wu2025plus}.
One-kernel-per-shape compilation recompiles the dynamic operators whenever new shapes are encountered, which suffers from the long compilation time of the backend compilers at the service time and the high compilation cache pressure due to the large number of operator shapes.
Multi-kernels refer to the solutions that compile multiple versions of the operator kernel before execution and dynamically select one for the encountered operator shape.
Some works require experts to write efficient kernels, e.g., cuBLAS~\cite{cublas} and PluS~\cite{wu2025plus}.
To prepare the kernel candidates automatically, the sampling-based methods~\cite{tvm_bucketing,zheng2022dietcode,zheng2023bladedisc} require obtaining the possible operator shapes to generate efficient kernels, while the sample-free methods~\cite{zhou2025sample} run bottom-up construction to get architecture-aligned kernels.
However, the limited kernel candidates are not flexible and general enough to support the complex model dynamism in practice, e.g., the various kinds of fused operators.

\textbf{Operator fusion for dynamic models.}
Both dynamic operator shapes and dynamic control flows need to be considered when making operator fusion decisions for dynamic models.
To deal with the dynamic shapes, some works~\cite{shen2021nimble,zhu2021disc} apply basic fusion for memory-intensive operators, while BladeDISC~\cite{zheng2023bladedisc} utilizes the equality relationship between tensor shapes, instead of the concrete shape values, to check locality information for advanced dynamic shape fusion decisions. %
To deal with the dynamic control flows, Tempo~\cite{silvestre2025tempo} focuses on recurrent computation graphs and eliminates the recurrent patterns via lifting.
DyCL~\cite{chen2023dycl} converts a dynamic neural network into multiple static sub-neural networks, each with no conditional statements and compiled independently.
Cocktailer~\cite{zhang2023cocktailer} considers three kinds of control flows, i.e., loops, branches, and recursion, and puts the control flow logic into kernels to process on the GPU, therefore reducing the CPU-accelerator synchronization overhead and enabling fusion opportunities (e.g., running independent operators in parallel) across control flow scopes.
However, these methods still cannot flexibly handle fusion opportunities in more complex control flows, e.g., fusing operators across multiple branches.
{\color{black}Lazy Tensors~\cite{suhan2021lazytensor} defers operation execution to accumulate a graph and then sends the accumulated graph to the XLA compiler~\cite{xla}. In this way, it can make use of the actual execution paths and operator shapes for compilation.
The accumulated graph is hashed to avoid unnecessary recompilation.
PyTorch 2~\cite{ansel2024pytorch} tries to reason the shapes to resolve control flow when given the model input and graph break when it fails.
However, both Lazy Tensors~\cite{suhan2021lazytensor} and PyTorch 2~\cite{ansel2024pytorch} suffer from time-consuming recompilation, and the hashed graph maintenance in Lazy Tensors can cause extra overhead.
By contrast, \ours~can deal with the actual execution paths and operator shapes at runtime efficiently without caching the fusion decisions or generated kernels, thanks to the virtual-machine-based efficient operator compiler.}

\section{Conclusion}
In this paper, we design a real-time compiler \ours, consisting of an operator compiler, which performs effective and efficient runtime operator compilation based on a virtual machine, and an operator fuser, which performs symbol-deduction-based fusion on static graphs and runtime fusion on dynamic graphs.
Both pattern- and stacking-based fusions are supported to increase fusion opportunities.
Evaluation on operators, subgraphs, and models shows that, compared with \tritonfull, \eagerfull~and \msfull, we are up
to {\color{black}11.77$\times$} better in terms of the operator/model efficiency and up to {\color{black}5} orders of magnitude faster in terms of the maximum compilation time, which validates our significant superiority in developing and deploying dynamic models.

\section*{Availability}
The code of \ours~is available at \url{https://gitcode.com/mindspore/dvm}.

\bibliographystyle{plain}
\bibliography{reference}

\end{document}